# Correlative analysis of structure and chemistry of Li$_x$FePO$_4$ platelets using 4D-STEM and X-ray ptychography

L.A. Hughes[1*], Benjamin H. Savitzky[1*], Haitao D. Deng[2*], Norman L. Jin[2], Eder G. Lomeli[2], Young-Sang Yu[3], David A. Shapiro[3], Patrick Herring[4], Abraham Anapolsky[4], William C. Chueh[2], Colin Ophus[1], and Andrew M. Minor[1,5]

[1.] National Center for Electron Microscopy, Molecular Foundry, Lawrence Berkeley National Laboratory, Berkeley, CA 94708, USA
[2.] Depart. of Mat. Sci. & Eng., Stanford University, Stanford, CA 94305, USA
[3.] Advanced Light Source, Lawrence Berkeley National Laboratory, Berkeley, CA 94720, USA
[4.] Toyota Research Institute, Los Altos, CA 94022, USA
[5.] Depart. of Mat. Sci. & Eng., University of California, Berkeley, CA 94720, USA
**\*these authors contributed equally**
**Corresponding authors: laurenhughes@gmail.com, aminor@lbl.gov, wchueh@stanford.edu**

## Abstract

Lithium iron phosphate (Li$_x$FePO$_4$), a cathode material used in rechargeable Li-ion batteries, phase separates upon de/lithiation under equilibrium. The interfacial structure and chemistry within these cathode materials affects Li-ion transport, and therefore battery performance. Correlative imaging of Li$_x$FePO$_4$ was performed using four-dimensional scanning transmission electron microscopy (4D-STEM), scanning transmission X-ray microscopy (STXM), and X-ray ptychography in order to analyze the local structure and chemistry of the same particle set. Over 50,000 diffraction patterns from 10 particles provided measurements of both structure and chemistry at a nanoscale spatial resolution (16.6-49.5 nm) over wide (several micron) fields-of-view with statistical robustness. Li$_x$FePO$_4$ particles at varying stages of delithiation were measured to examine the evolution of structure and chemistry as a function of delithiation. In lithiated and delithiated particles, local variations were observed in the degree of lithiation even while local lattice structures remained comparatively constant, and calculation of linear coefficients of chemical expansion suggest pinning of the lattice structures in these populations. Partially delithiated particles displayed broadly core-shell-like structures, however, with highly variable behavior both locally and per individual particle that exhibited distinctive intermediate regions at the interface between phases, and pockets within the lithiated core that correspond to FePO$_4$ in structure and chemistry. The results provide insight into the Li$_x$FePO$_4$ system, subtleties in the scope and applicability of Vegard's law (linear lattice parameter-composition behavior) under local versus global measurements, and demonstrate a powerful new combination of experimental and analytical modalities for bridging the crucial gap between local and statistical characterization.

## Introduction

Lithium iron phosphate (Li$_x$FePO$_4$) is a viable commercial cathode material for lithium (Li) batteries used in portable electronic devices and zero-emission vehicles due its high rate performance, good cycle life, and low cost.[1,2] However, mechanisms that affect Li-ion transport, which plays a dominant role in rate performance and cycle life, are not fully understood in Li$_x$FePO$_4$.[3–5] A key question is how exactly the structure and chemistry of Li$_x$FePO$_4$ change as it phase separates into Li-rich (LiFePO$_4$) and Li-poor (FePO$_4$) phase domains.[6–8] To fully understand the functional behavior of Li$_x$FePO$_4$, the chemical reactions and phase transformations that occur upon de/lithiation must be clearly determined.

A multitude of microscopy techniques have been applied to analyze the Li distribution and crystallographic information of Li$_x$FePO$_4$ particles in order to determine the insertion/desertion mechanisms during cycling.[9] Scanning transmission X-ray microscopy (STXM) as well as X-ray ptychography have previously shown the distribution of Li during de/lithiation.[10] Chueh et al. show the effect of Li surface diffusion on phase transformation behavior of Li$_x$FePO$_4$ platelets via STXM.[8] They also show the incoherent nanodomains within Li$_x$FePO$_4$ ellipsoidal nanoparticles behave as distinct particles during delithiation via X-ray ptychography.[11] High resolution transmission electron microscopy (HRTEM) and selected area electron diffraction (SAED) have isolated defect concentration and crystallographic information (lattice parameters, strain, orientation) in LiFePO$_4$ and FePO$_4$.[12–14] Electron energy loss spectroscopy (EELS) of the Fe-L$_{2,3}$ and O-K edges, as well as Li-K edge using low loss EELS, have revealed variations in electronic bonding configuration and Li concentration at the interface in which these two phases interact.[15–17] Annular bright field (ABF) imaging has captured changes in local atomic structure and Li-ion location at the phase separated interface within Li$_x$FePO$_4$ platelets.[18,19] While these microscopy techniques successfully highlight structural and chemical changes at the phase separated interfaces within de/lithiated Li$_x$FePO$_4$, they did not correlate both structure and chemistry globally and locally within the same particles as their field of view was limited to selected regions of interest. Restrictions within these characterization techniques, due to sample thickness and statistics, have led to discrepancies between the experimental observations and the proposed theoretical models regarding the mechanisms of de/lithiation within Li$_x$FePO$_4$.[17,20]

In regards to analyzing structural effects, automated crystal orientation mapping (ACOM) in TEM in combination with energy filtered TEM (EFTEM) has also been used to analyze phase transformation behavior in LiFePO$_4$, providing some statistical information on crystal orientation and structure.[21,22] For this technique, the experimental electron diffraction patterns are compared to a database of simulated diffraction templates calculated from literature. LiFePO$_4$ and FePO$_4$ are both orthorhombic with space group *Pnma*[13,23] and reported crystalline lattice parameters of a = 10.33 Å, b = 6.01 Å, c= 4.69 Å for LiFePO$_4$, and a =9.81 Å, b=5.79 Å, c=4.78 Å for FePO$_4$, respectively.[13,23] However, the crystal structure and lattice parameters for potential intermediate phases are unknown due to phase separation; simulated datasets must then make assumptions regarding any possible intermediate phases. For example, an intermediate phase within cycled Li$_x$FePO$_4$ is assumed to be a solid solution of LiFePO$_4$ and FePO$_4$ in which 50% of the sites are occupied by Li and the lattice parameters are an average of the lattice parameters of LiFePO$_4$ and FePO$_4$ phases.[22] Phase mapping over micrometers at nanometer resolution is therefore possible with ACOM-TEM, but the assumptions required to make the method viable may obscure novel or unexpected structures within de/lithiated Li$_x$FePO$_4$ such as new phases or any non-linearity in lattice parameter variation with composition.

In previous work, the study of phase distribution within Li$_x$FePO$_4$ as a function of de/lithiation has been performed separately via diffraction, spectroscopy, or X-ray techniques.[11,24,25] However, to accurately delineate the structural and chemical transformations as well as their relationship to one another, the data acquired from diffraction and spectroscopy techniques must be directly comparable and statistically applicable.[21,26–28] Therefore, in this correlative study, four-dimensional scanning transmission electron microscopy (4D-STEM) and two X-ray microscopy techniques, STXM and X-ray ptychography, were performed on the same $Li_xFePO_4$ platelets. This combination allows for pixel-by-pixel correlation of the datasets, such that the lattice parameters and lithium content can be extracted and compared over fields-of-view, which span complete particles with statistical robustness, and was not previously achievable due to technique and data acquisition limitations. In our study, ten particles were examined – three lithiated, five partially delithiated, and two delithiated – comprising over 50,000 diffraction patterns. Due to data-size and electron beam damage constraints, all ten particles were initially measured by 4D-STEM with a comparatively large step-size of 49.5 nm between collected diffraction patterns. A single partially delithiated particle was then examined in finer detail with a step-size of 16.6 nm. In each case, the electron and X-ray data were used to extract the lattice parameters and percent lithiation, and the results were correlated to allow for direct, local comparison. The results show local variations, which likely would be lost under either broad-beam characterization or traditional high resolution, but narrow field, microscopy. In a separate publication using this data, inverse learning of the chemo-mechanics is performed and a variety of models are subsequently analyzed.[29] Thus, the acquired correlative imaging results demonstrate the capabilities of 4D-STEM combined with STXM/X-ray ptychography for isolation and identification of structural variations, which is a necessary component for the development of insertion/desertion kinetic modeling of $Li_xFePO_4$.

Within this work, the $Li_xFePO_4$ particles are categorized by three chemical ranges based on the average Li composition (x) of each particle: $0 \leq x \leq 0.15$ is defined as FP, $0.15 < x \leq 0.85$ is defined as $L_xFP$, $0.85 < x \leq 1$ is defined as LFP. When discussing literature values or comparing to the 'pure' bulk phase, $LiFePO_4$ or $FePO_4$ are used. $Li_xFePO_4$ is a catchall for all compositions.

**Results**

**Pixel-by-pixel correlation of $Li_xFePO_4$ particles using 4D-STEM and X-ray microscopy**

Figure 1 shows a- and c-lattice parameter maps and chemical composition maps for nine $Li_xPO_4$ platelets. For these maps, red denotes values of each parameter (a- and c- lattice parameters as well as chemical composition) corresponding to $FePO_4$; yellow denotes values corresponding to $LiFePO_4$; and blue denotes intermediate values. These plots were produced by first performing 4D-STEM with a 49.5 nm step-size and then performing STXM or X-ray ptychography on the exact same $Li_xFePO_4$ particles, with each platelet aligned along the [010] direction. A single crystallographic orientation was observed for each platelet. Lattice parameters were extracted by fitting the Bragg scattering within each 4D-STEM diffraction pattern (see Methods and Supplemental Material).[30] Chemical composition values were determined using the X-ray signal obtained from 5 different energy slits about the Fe-K edge (Methods and Supplemental Material). Correlative analysis produced additional information such as HAADF-STEM, X-ray optical density, and shear and rotational strain for each particle. Pixel-by-pixel correlation between the X-ray and 4D-STEM signals was performed using an affine transformation optimized by gradient descent (see Methods and Supplemental Material). Sixteen distinct image-like information channels were ultimately extracted for each particle (see Supplemental Material). While we focus primarily on the a-lattice, c-lattice, and percent lithiation channels in this work, this data is highly information rich – for instance, the HAADF-STEM and X-ray optical density images for all the $Li_xFePO_4$ particles show the presence of internal voids of varying size and shape. We therefore made this data freely and publicly available for further study – see Data Availability.

From the Li-composition-distribution maps, LFP platelets are shown to be fully lithiated with minimal fluctuation in Li content regardless of internal void structure (Particles 1, 2, and 3 in Figure 1). We measure both a- and c-lattice parameter values that are consistent with the literature, finding a = 10.33 ± 0.13 Å and c = 4.69 ± 0.05 Å, where the error terms quoted here represents the standard deviation of the distribution. The noisier appearance of the c-lattice parameter maps in Fig. 1 compared to the a-lattice parameter is likely due to the fact that, for the a-lattice parameter, the two peaks of bimodal distribution are cleanly separated; while for the c-lattice parameter, the two peaks of bimodal distribution are not distinctly separated. This difference in bimodal distribution is discussed further in the analysis of the singular $L_xFP$ particle within the next section. Overall, the experimental LFP platelets, from 4D-STEM and STXM analysis, demonstrate lattice parameters and chemistry that correspond well with literature.[1]

The Li composition-distribution maps of FP platelets, in contrast, demonstrate that these platelets are not fully delithiated. Particle 5 shows a modest fluctuation of Li content, resulting in purple-blue regions throughout the overall platelet; while Particle 4 shows uniform, minimal Li content (Figure 1). Interestingly, the a-lattice parameter map for Particle 5 exhibits small, concentrated clusters with intermediate a-lattice parameter values, approximately 10.0 Å. These clusters overlap with fluctuations in the Li concentration. However, the variation within the Li concentration is more diffuse throughout the FP platelet as the platelet shows a Li concentration gradient change from the fully delithiated regions to the intermediate lithiated regions, whereas the intermediate a-lattice parameter values are distinct from the fully delithiated regions, i.e., color gradation within the a-lattice parameter map for Particle 5 does not directly correspond to the color gradation of the Li composition map. The c-lattice parameter maps for FP, similar to LFP particles, show intermediate lattice parameter values, which fall within the literature range.[1,13] Modest fluctuation of Li within delithiated FP can be attributed to a number of factors related to defect concentration, surface morphology, and lithiation pathways that are all sample dependent.[31]

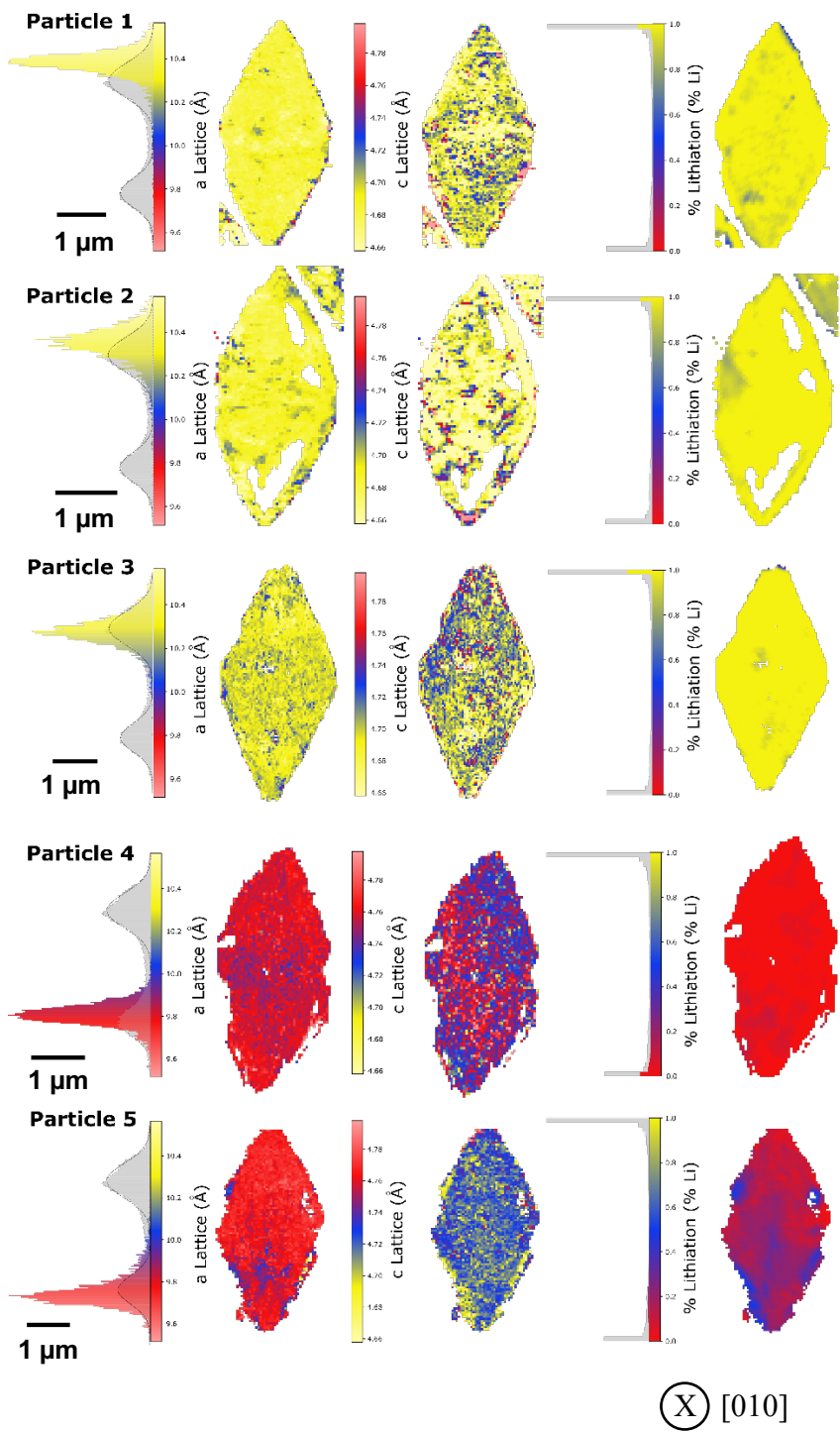

**Figure 1. Pixel-by-pixel correlation of $Li_xFePO_4$ particles:** The a-lattice parameter map, the c-lattice parameter map, and % lithiation map for LFP (Particles 1-3), FP (Particles 4-5), $L_xFP$ (Particles 6-9). The color scheme is the same: red denotes regions that have chemistry or lattice parameters that correspond to $FePO_4$, yellow denotes regions that have chemistry or lattice parameters that correspond to $LiFePO_4$, blue denotes regions that have chemistry and lattice parameters that correspond to their mean. The lithium composition was determined using STXM for LFP and FP particles, and X-ray ptychography for $L_xFP$. The step-size used for the 4D-STEM measurements was 49.5 nm. Definition of the colormaps is discussed further in the Supplemental Materials.

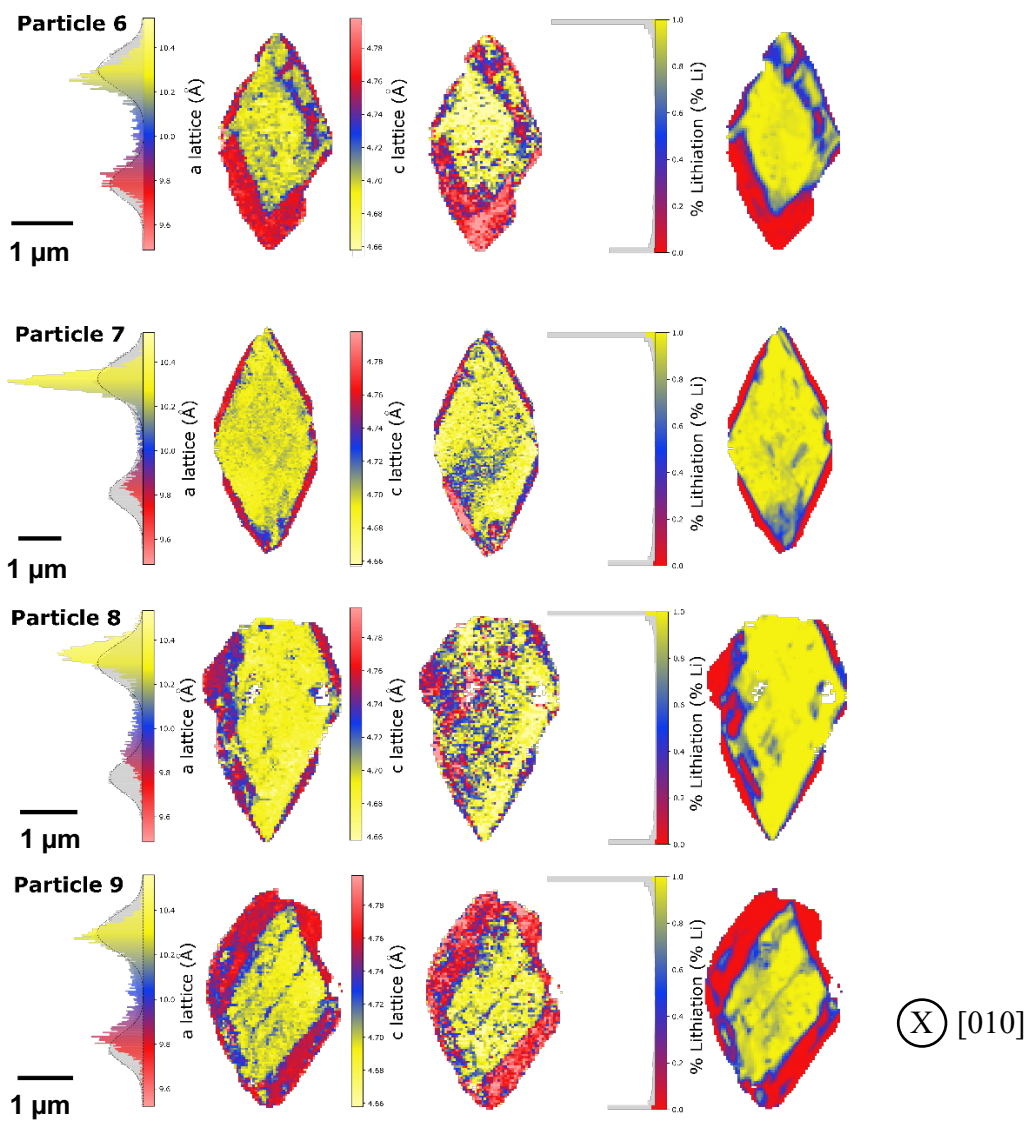

The L$_x$FP platelets, overall, show a Li-rich core and Li-poor outer shell (Particles 6-9 in Figure 1). In the delithiated (lithiated) regions, we find an a-lattice parameter of 9.82 ± 0.16 Å (10.22 ± 0.15 Å) using mean statistics, and a value of 9.76 ± 0.16 Å (10.28 ± 0.15 Å) with median statistics, where in both cases the error term is the standard deviation. Similarly, for the c-lattice parameter we find 4.76 ± 0.03 Å (4.70 ± 0.03 Å) in the means and 4.77 ± 0.03 Å (4.69 ± 0.03 Å) in the medians. In terms of percentage contraction/expansion from the lithiated to the delithiated regions, the a-lattice is found to contract by 3.9% in the means and 5.1% in the medians. The c-lattice is found to expand by 1.3% in the means and 1.7% in the medians. The mean and median statistics vary meaningfully here due to distributions with meaningful skewness, and the full distributions can be found in the Supplemental Material. In comparison to the values reported by Zhang when Li-ions were extracted from Li$_x$FePO$_4$, we find similar c-expansion values to theirs (1.9%), and our a-contraction values are somewhat less than their report (6.77%).[1]

The interface between the Li-rich core and Li-poor shell of the L$_x$FP platelets show intermediate a-lattice parameter values and Li composition, which is highlighted in blue within these maps. Each $L_xFP$ particle (Particles 6-9) exhibits a different, non-uniform core-shell behavior, in which there is not a stark separation of the Li-rich and Li-poor domains, e.g., there are Li-poor pockets within the lithiated core and there are partially lithiated pockets within the delithiated outer shell. Some partially lithiated and fully delithiated pockets are correlated with the presence of voids. For instance, a partially delithiated region, denoted in blue, at the base of $L_xFP$ Particle 7 follows the shape of the connected void structures observed in both the HAADF-STEM and X-ray optical density images shown in the Supplemental Material. However, a clear relationship between composition and void structure is not observed for all the particles. We explore this relationship between local structure and Li chemistry in a following section.

**Structure-composition relationship for all $Li_xPO_4$ particles**

Vegard's law predicts that in a solid solution of two materials, the lattice parameters will scale linearly with the weighted mean of those two constituents, i.e., linear lattice parameter-composition behavior is observed within a uniform, stress-free state. In Figure 2, the a-lattice parameter as a function of lithiation for all the LFP, $L_xFP$, and FP particles is plotted. For the LFP and the FP particles, the a-lattice parameter is shown to be consistent with bulk phase $LiFePO_4$ or $FePO_4$, respectively, regardless of changes in lithiation (Figure 2a and 2c). This pinning of the a-lattice parameter indicates the mean behavior of these particles does not always follow Vegard's law. Mechanistically, this behavior corresponds with the population density curves (dashed lines), which show the vast majority of counts are fully lithiated/delithiated in the LFP/FP, dropping rapidly for lithiations below 90% in LFP and above 10% in FP. We therefore surmise this pinning of the a-lattice parameter with respect to % lithiation is caused by the predominantly 'pure' LFP or FP crystal structures preventing lattice relaxation in the relatively small and few regions of partial lithiation.

In contrast, Figure 2b shows Vegard's law is, on average, obeyed in the $L_xFP$ particles. The mean of the a-lattice parameter is shown to vary linearly with the Li content. We reiterate, however, that these measurements are spatially averaged, which masks any local variability and is therefore a significant qualifier when drawing conclusions regarding structure-composition behavior of de/lithiated $Li_xFePO_4$.

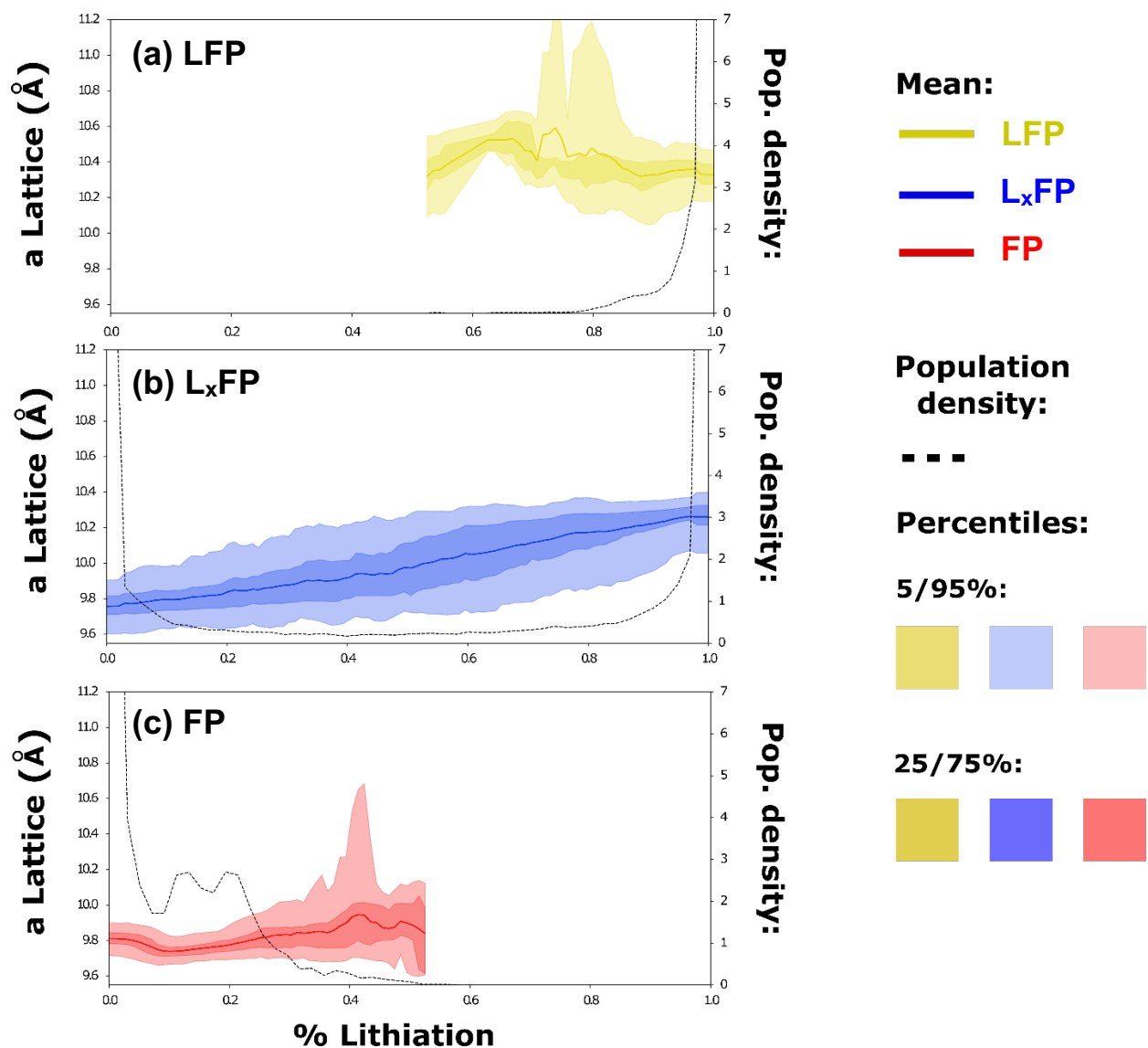

**Figure 2. Population density, mean, and percentile for all $Li_xFePO_4$ particles:** Changes in the a-lattice parameter compared to % lithiation for (a) all LFP particles, (b) all $L_x$FP particles, and (c) all FP particles.

To quantify this analysis, we computed the linear coefficients of chemical expansion, shown in Table 1. We computed linear fits to the a- and c- lattice parameters versus lithiation distributions for each particle, then found the mean and standard deviations above using all particles at a given lithiation. See Supplemental Information for more information

| Type | $\alpha_a$ (Å/%Li) | $\alpha_c$ (Å/%Li) |
|---|---|---|
| LFP | −0.033 ± 0.219 | 0.062 ± 0.079 |
| L$_x$FP | 0.502 ± 0.025 | −0.073 ± 0.016 |
| FP | 0.174 ± 0.291 | −0.046 ± 0.017 |
| Lit. | 0.524 | −0.089 |

**Table 1. Chemical Expansion Coefficients of L$_x$FP**: The linear coefficients describing the statistical expansion of the a- and c- lattice parameters ($\alpha_a$ and $\alpha_c$) with lithiation.

**Detailed analysis of a single L$_x$FP particle**

A higher-resolution 4D-STEM dataset of L$_x$FP was also acquired with a smaller step-size of 16.6 nm (Particle 10, Figure 3). In Figure 3, pixel-by-pixel correlated maps of the a- and c- lattice parameters and Li composition were generated via 4D-STEM and X-ray ptychography. The phase separation interface between the Li-rich core and the Li-poor shell of the 16.6 nm step-size map shows intermediate a-lattice parameters and Li chemical compositions, denoted in blue, similar to the interfaces observed in the 49.5 nm step-size L$_x$FP maps (Particles 6-9, Figure 1).

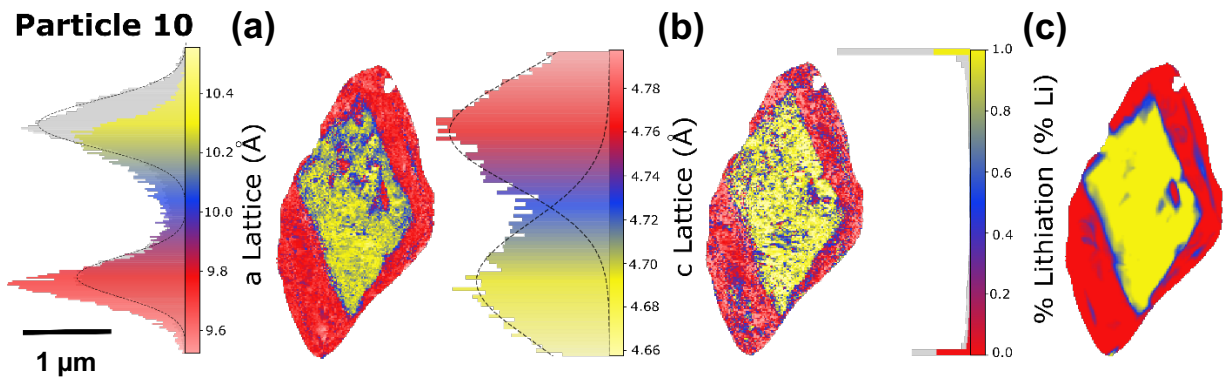

**Figure 3. Pixel-by-pixel correlation of L$_x$FP:** (a) a-lattice parameter map, (b) the c-lattice parameter map, (c) % lithiation map. For lattice parameter and lithium composition-distribution maps, the color scheme is the same as Figure 1. The a- and c-lattice parameter and % lithiation maps also have corresponding histograms in which the grey-scale histogram corresponds to all the data taken from the ten particles and the color-scale histogram corresponds the data acquired from the individual particle.

The histograms associated with each map show the populations of the relevant platelet quantities. The grey-scale histograms are populations derived from the data of all ten particles, while the color-scale histograms use the data of only Particle 10. The dashed lines show the pair of best-fit gaussians to the whole-population distributions. If the fit gaussians are taken as representative of the "pure" phase parameter distributions, it is apparent the a-lattice parameter histogram consists of two distinct, well segmented lobes with non-overlapping tails. Yet, this particle contains a statistically significant number of counts in between the two lobes and above the gaussian tails, and in the associated map it is apparent these counts arise primarily at and near the phase separation interfaces (blue regions). Thus, there exists narrow, but well-defined regions, which are structurally distinct from the pure phases. In contrast, the histogram of the c-lattice parameter is bimodal, but there is significant overlap between the lobes of the distribution associated with the fully lithiated and fully delithiated regions (Figure 3b). This phenomenology is related to the comparatively small median c-lattice parameter expansion of 1.7% compared to the 5.1% contraction of the a-lattice parameter, which causes difficulty in statistically distinguishing a separation between the two phases by the c-lattice parameter values as noted elsewhere in the literature.[13] We also note that this overlap between the normal distribution of the LFP and FP regions for the c-lattice parameter, and the comparative size scales spanned by the color bars for the a- and c-lattice parameters, are likely the cause of the signal to noise difference between the two structure maps.

In the analysis of $L_x$FP for Figure 2b, the a-lattice parameters were found to scale linearly with Li content after spatial averaging and a perspective of the overall structure-chemistry behavior was gained for these particles. However, 4D-STEM in combination with X-ray ptychography allows for analysis of the fine structural variations along with the average bulk variations within a particle. From the perspective of Figure 3, there are two interesting local structure-chemistry features. Firstly, within the Li-rich core of the $L_x$FP Particle 10, there is a contained pocket in which the a- and c-lattice parameters and Li content corresponds to the FePO$_4$ phase. This delithiated pocket is consistent with the location of a void observed in the HAADF-STEM and X-ray optical density images (see Methods and Supplemental Material). Secondly, within the Li-poor shell of the $L_x$FP Particle 10, there are regions which show intermediate a- and c- lattice parameters with partial delithiation. These intermediate regions mimic the line and shape of the phase separation interface and may thereby indicate the movement of the phase separation interface within the particle during the delithiation process.

**Error analysis and statistics in $L_x$FP Particle 10**

The particles examined were ~300-400 nm thick and non-uniform in thickness. This sample thickness creates significant challenges in analysis via the electron beam due to multiple scattering effects.[32] Thinner samples are preferable from a characterization perspective, though not always possible with real-world samples. The size of particles can alter their qualitative behaviors; and, in this instance, platelets with the best figure of merit for function, such as electrochemical cycling, are thick to an electron beam. In these cases, error analysis is essential to understand the realm of validity of any quantitative analysis.

Here we performed cross validation to calculate a root mean square error (RMSE) associated with the fit lattice vectors at each scan position. Briefly, this process involved fitting the lattice vectors to a subset of the detected Bragg peaks at each scan position, then computing how much the remaining subset of peaks deviated from their predicted position given these lattice vectors. The result is a pixelwise map of RMSE error, shown in Figure 4a. More details, and discussion of why this approach is preferable to quoting the lattice fitting error, are in Supplemental Material. We found the majority of the lattice vectors fits to be trustworthy, but there are several regions of high RMSE in which results are suspect. Thresholding the RMSE image to generate a mask gives the green region within Figure 4e. Examining the diffraction patterns provides insight into why the error is increased within these regions. Figure 4m shows the average of 30 diffraction patterns from a selected high RMSE region (shown in green in Figure 4f), which has an extremely low signal-to-noise ratio and no diffraction spots are visible to the eye. In general, we find the high RMSE regions fall into three categories: (1) very low SNR, (2) most detected Bragg peaks are colinear, or (3) redistribution of intensity inside the disks near the particle edges. The low SNR diffraction patterns typically appear to be due to sample thickness, additional material agglomerated on the particle, or local lattice tilt. See the Supplemental Material for further discussion. Most crucially, careful assessment of the CV error reveals that, in spite of the large sample thickness, the fit lattice vectors are trustworthy for the vast majority of pixels, suggesting the validity of the subsequent analysis.

We also computed local deviation from Vegard's law. Our metric δ is given by $\delta = a - a_{FP} - \%Li(a_{LFP} - a_{FP})$ and represents the deviation, in units of length, of the local a-lattice parameter from the value predicted by Vegard's law. In Particle 10, we found several pockets of anomalous contraction of a-lattice parameter in the upper area of the lithiated core, colored in red in Figure 4b. Comparing this contraction directly to the measured Li content (Figure 4c and 4e), there is no local change in Li and the regions are nominally fully lithiated. To understand if this anomalous contraction represents a real change to the lattice, we then isolated these regions, using a threshold to generate a mask, shown in blue in Figure 4e. An average of 30 diffraction patterns from this region is shown in Figure 4n. We found that, while these regions had a sufficiently high SNR to detect Bragg scattering, the disks themselves were smeared, indicating possibly a tilt of the lattice or overlapping structures, preventing conclusive measurement of the a-lattice parameter in these regions. In light of these considerations we estimated the systematic error associated with our measured lattice parameters and found a systematic error associated with the a-lattice of 3.6 ± 1.5 pm and with the c-lattice of 2.2 ± 1.8 pm. See the Supplemental Materials.

Using a cutoff at 50% lithiation, we segmented the remaining diffraction patterns into regions classified as FP and LFP, shown in red and yellow respectively in Figure 4e. With these regions, we can then compute structural statistics based on local behavior while remaining confident in the validity of the measurements. Figures 4k and 4l show images averaged over 30 diffraction patterns each from the FP and LFP regions shown in Figure 4f. Circular Bragg disks with good SNR are visible by eye. Figures 4g-4j shows histograms of several parameters of interest for the two phases: the a- and c- lattice parameters, the percent lithiation, and the optical density (a good proxy for thickness). The a- and c- lattice values show means consistent with the literature. The overlap of the tails of the distributions in the c-lattice parameter indicate that local variation is on similar order to the difference between $c_{FP}$ and $c_{LFP}$, making the c-lattice value a poor metric for phase determination. The optical density tends to be higher in the LFP, which is sensible given the tendency of platelets to taper at their edges where the FP is located. Careful error analysis and masking of suspect diffraction patterns makes it possible to make these measurements with confidence.

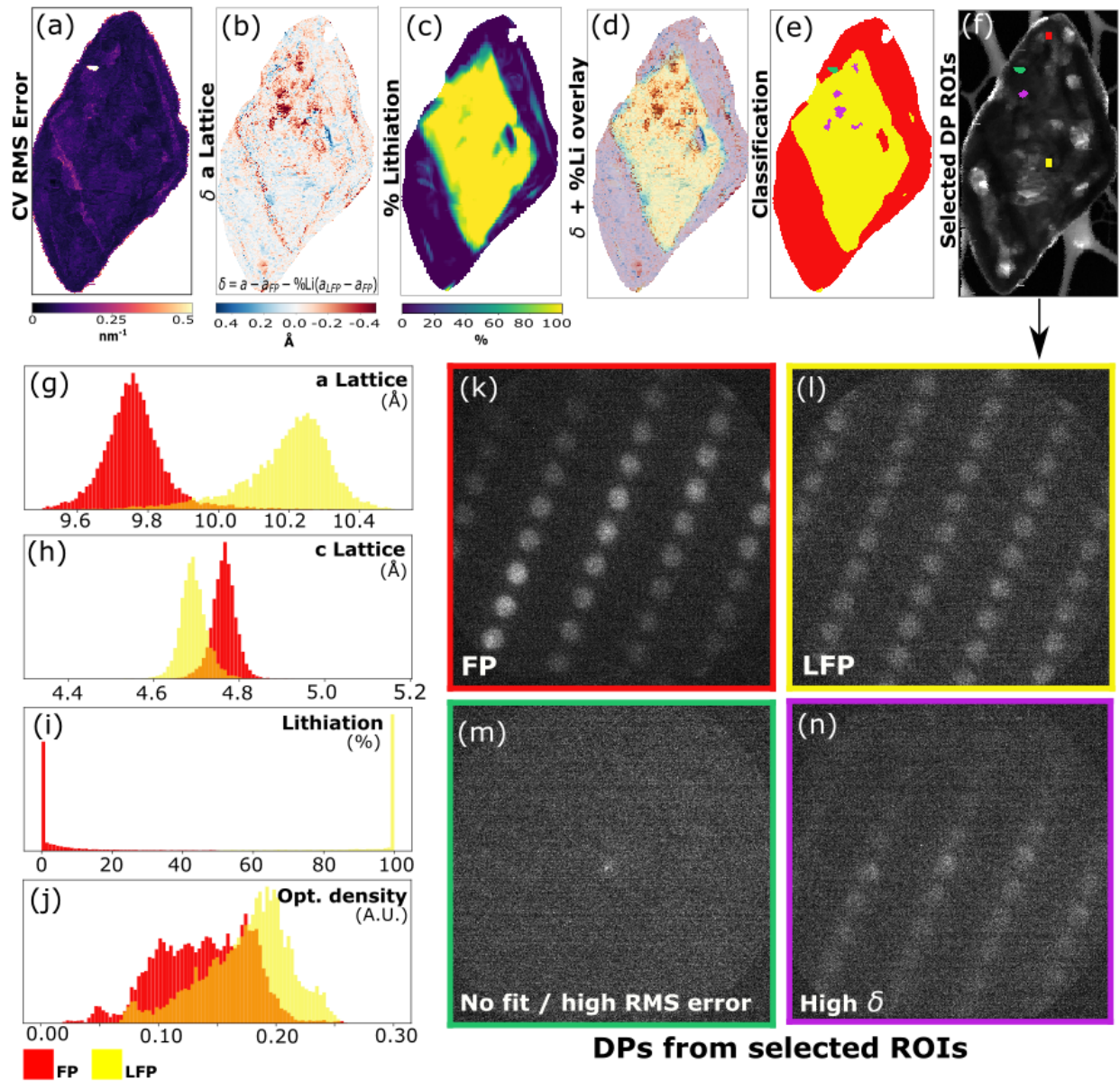

**Figure 4. Error analysis and segmentation of $L_xFP$ Particle 10**: (a) A map showing cross validation root mean square error associated with the measured lattice vectors at each pixel; (b) the deviation in the a-lattice parameter at each pixel relative to the values expected using Vegard's law in combination with the experimental, bulk FP and LFP values; (c) % lithiation of $L_xFP$ Particle 10; (d) an overlay of the changes in a-lattice parameter at each pixel and the % lithiation map; (e) a segmentation of regions of statistically significant a-lattice parameter deviation from Vegard's law (blue) and RMS error (green) as compared to bulk LFP (yellow) and FP (red); (f) ADF image with colorized size-limited regions of these four segments (FP, LFP, deviation of Vegard's, and RMS error with their corresponding diffraction patterns (k-n)). Histograms of the (g) a-lattice parameter, (h) c-lattice parameter, (i) % lithiation and (j) thickness values of $Li_{0.5}FePO_4$ Particle 10, where red/yellow indicate FP/LFP, and overlap between the two histograms curves is in orange.

**Discussion**

Excluding Particle 7, which has sharply defined edges and is the closest to a diamond shape, the remaining $L_x$FP platelets exhibit delithiated shells that are non-uniform (Particles 6-9, Figure 1). Certain particle edges of the $L_x$FP delithiated more rapidly compared to other particle edges. This variation is likely due to changes in the surface energy of the linear edge compared to the rough or curved edge of these platelets, which then alters the rate of Li-ion diffusion.[13,33,34] The non-uniform platelet delithiation, the distinct and unevenly distributed phase-separation interfaces demonstrate the importance of these nanoscale resolution, micrometer-scale field-of-view, and statistically robust datasets. As we demonstrated within this work, the $L_x$FP particles show changes in structure and chemistry across a few nanometers distance as well as between particles.

Combining 4D-STEM and X-ray microscopy techniques allows for the multiscale analysis necessary for observing the structural and chemical changes in which $Li_x FePO_4$ and other battery materials undergo as a function of delithiation. This work exhibits that $Li_x FePO_4$ platelets have a-lattice parameters values which do not follow the Vegard's law, i.e., linear lattice parameter-composition behavior, even under position averaging, in the FP and LFP states (Figure 2); and that the $L_x$FP platelets show a statistically significant fraction of pixels that correspond to the intermediate values, which varies locally within particles and from particle to particle (Figure 1 and 3). Thus, the results illustrate the significant variability of structure and chemistry within these particles and between differing particles, which undoubtedly affects analysis when data acquisition is confined to a narrow scope. Data acquisition which is confined solely to a narrow field of view (often associated with electron microscopy experiments) or a wide field of view but spatially averaged (often associated with X-ray microscopy experiments) can miss or overlook key components of analysis crucial to fundamentally understanding of battery materials.[3,4,21]

In total, the datasets presented combined local nanoscale resolution across several micrometer fields of view, correlated structural and chemical channels, and the statistical robustness afforded by new "big data" experimental methods like 4D-STEM. With no assumptions regarding ascribed literature values or diffraction pattern templates, we calculated the lattice parameters and compositions directly from the raw data. The unique combination of local and statistical measurements enabled a more nuanced look into Vegard's law in this system: for fully lithiated/delithiated particles, a spatially-averaged description (Figure 2) indicated deviations from Vegard's law, even while it was obeyed within the vast majority of local measurements (Figure 1). In partially delithiated particles, Vegard's law was obeyed under spatial averaging (Figure 2), while the local picture was more complex and not yet fully clear. These local and statistical measurements also identified intermediate structure-chemistry values at the phase separation interface, i.e. a- and c- lattice parameters as well as Li content in between the $FePO_4$ and $LiFePO_4$ phases (Figure 3 and 4).

**Conclusion**

Ten $Li_x FePO_4$ platelets at different stages of chemical delithiation (lithiated LFP, $L_x$FP, and delithiated FP) were characterized using 4D-STEM, STXM, and X-ray ptychography. Our correlative microscopy approach allows for a pixel-by-pixel correlation and comparison of the structure and chemistry of Li$_x$FePO$_4$. From the acquired 4D-STEM and X-ray microscopy datasets, the a- and c- lattice parameter values and the Li concentration were calculated to determine the average and local structure-chemistry relationship. We observe that, on average, the L$_x$FP platelets follow Vegard's law, in which the changes in structure and chemistry are directly correlated. Thus, this correlative technique using 4D-STEM and X-ray microscopy independently calculates the structure and chemistry of µm-size platelets at nm-resolution from raw data and provides a methodology for sample analysis that encompasses robust statistics, high spatial resolution, a particle-size field of view, and minimal assumptions regarding material characteristics and properties.

## Acknowledgments

This project was supported by the Toyota Research Institute. Work at the Molecular Foundry and the Advanced Light Source was supported by the Office of Science, Office of Basic Energy Sciences, of the U.S. Department of Energy under Contract No. DE-AC02-05CH11231.

## Contributions:

H.D.D., N.J., and E.L. synthesized and prepared the samples. H.D.D. and N.J. performed STXM and ptychography experiments. Y.-S.Y. and D.A.S. assisted in the STXM and ptychography experiments. H.D.D. analyzed the STXM and X-ray spectro-ptychography data. L.A.H. performed the 4D-STEM experiments. L.A.H. and B.H.S. analyzed 4D-STEM data. C.O. performed the image registration. L.A.H. prepared the initial draft of the manuscript. All authors contributed to the discussion of the results and writing of the manuscript. W.C and A.M.M supervised the project.

## Data availability:

The 4D-STEM and X-ray microscopy data associated with this manuscript can be found at: https://data.matr.io/6/.

**Methods and Supplemental** for L. Hughes, et al. "Correlate Correlative analysis of structure and chemistry of $Li_xFePO_4$ platelets using 4D-STEM and X-ray ptychography"

### Synthesis and delithiation of $Li_xFePO_4$ platelets

$Li_xFePO_4$ platelets were synthesized using a solvothermal method. All precursors were purchased from Sigma-Aldrich. In detail, 6 mL of 1M $H_3PO_4$ was mixed with 24 mL of polyethylene glycol 400. Afterwards, 18 mL of 1M LiOH(aq) was added to precipitate $Li_3PO_4$. The mixture was constantly bubbled with a flow rate of ~ 50 mL/min with dry $N_2$ for ~16 hours to deoxygenate. Next, 12 mL of deoxygenated $H_2O$ was added via a Schlenk line to $FeSO_4 \cdot 7H_2O$ powder, which was pre-dried under vacuum. The $FeSO_4$ solution was then injected to the $Li_3PO_4$ suspension without oxygen exposure, and the entire mixture was transferred to a 100-mL teflon lined autoclave. The autoclave was heated to 140 ºC for 1 h, then to 210 ºC for 17 h. We note that the particles from this synthesis produces smooth, flat and well-faceted platelet particles, but not void free.

Fully lithiated LFP particles were chemically delithiated to $L_xFP$ (x = 0.5, chemical composition) and FP to avoid the inter-particle phase separation, also known as mosaic pattern. In detail, pristine LFP particles through a redox titration platform. This ensures the formation of interfaces after phase separation, and eliminates the kinetic heterogeneities due to carbon coating during electrochemical delithiation. Dilute 50 mL of $H_2O_2$(aq) was stoichiometrically added to 50mL LFP particle suspension (1mg/mL) to obtain $L_xFP$ and FP. To minimize the morphological damage to the particles due to fast non-uniform local reaction, titration rate was optimized to be 5 mL/h, and the suspension was constantly stirred at 500 rpm.

### 4D-STEM characterization and py4DSTEM

4D-STEM is a microscopy technique in which a condensed electron beam is rastered across a sample surface (Figure 1S). At each scan position, a convergent beam electron diffraction (CBED) pattern is acquired and these patterns are amassed to form a 4D-STEM dataset. 4D-STEM datasets of fully lithiated LFP, $L_xFP$, and fully delithiated FP platelets were obtained using a TitanX, operating at an accelerating voltage of 300 kV. For data acquisition, the convergence semi-angle was 0.48 mRAD with a 40 μm condenser (C2), spot size of 8, and camera length of 600 mm. The probe size was in the range of 2.2-2.5 nm with scans acquired at step sizes of 16.6 nm and 49.5 nm. The particles were all oriented along the [010] direction for 4D-STEM measurements.

Lattice parameter maps were generated from the 4D-STEM datasets using py4DSTEM, an open-source python program available on Github (https://github.com/py4dstem). Using py4DSTEM, data analysis is firstly focused on identifying the Bragg discs and their location within each diffraction pattern (Figure 2S). This identification is performed using a vacuum probe template, i.e., an image or image stack of the probe in vacuum with the experimental microscope parameters, cross correlated with each individual diffraction pattern (Figure 2Sa). This calculated cross correlation locates the position of the Bragg disks within all diffraction patterns and a Bragg vector map is then formed (Figure 2Sb). From this Bragg vector map, the vector length and angles are calculated, applied to all diffraction patterns, and strain maps are generated (Figure 2Sd, e). Further processing is done to calculate the lattice parameter maps from the generated strain maps using an aluminum standard 4D-STEM dataset to correct for elliptical distortion and provide pixel calibration (Figure 2Sb, c). Finally, Figure 2Sf shows the rotational calibration performed on the data analysis to ensure the real space HAADF-STEM images and the reciprocal diffraction space images are correlated.

**(a) Vacuum probe, find Bragg discs**

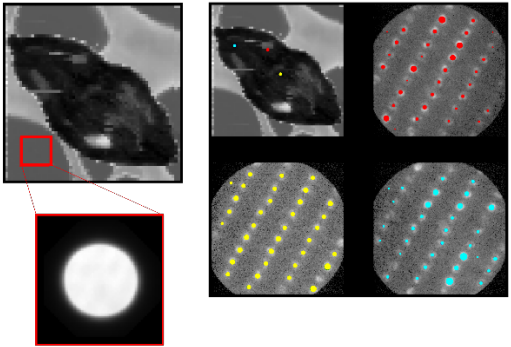

**(b) Calibration: diffraction shifts**

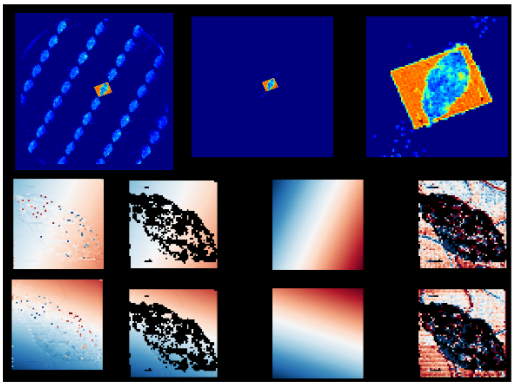

**(c) Calibration: elliptical distortion**

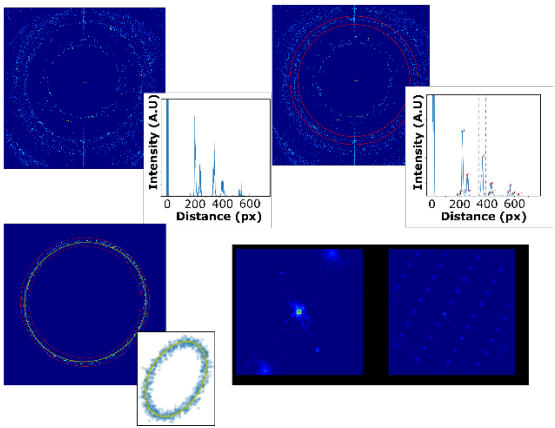

**(d) Calculate and fit lattice vectors**

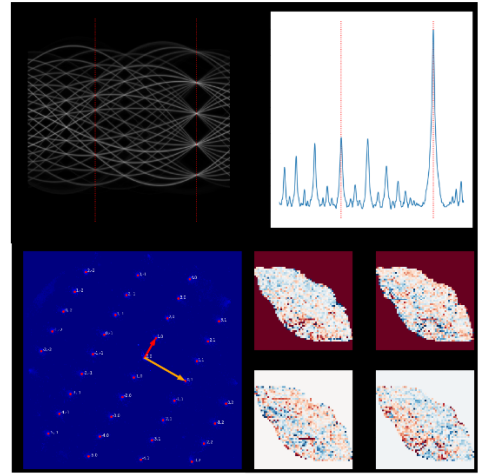

**(e) Generate strain maps**

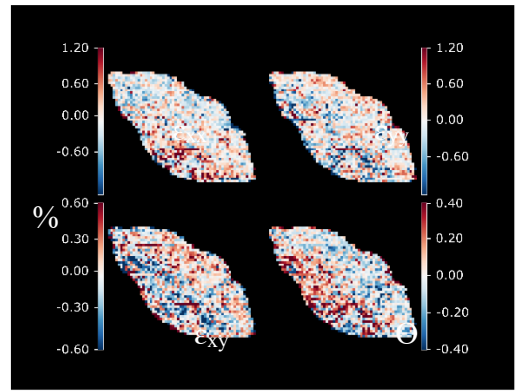

**(f) Calibration: rotation**

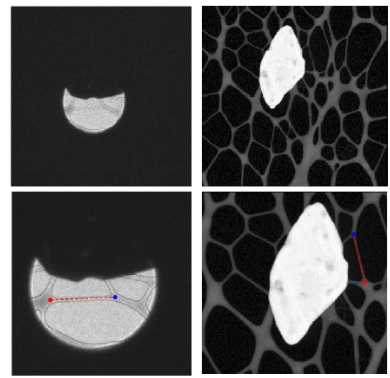

**Figure 2S.** Workflow example of py4DSTEM. (a) Vacuum probe template, (b) Bragg vector map generated from diffraction patterns and subsequent beam shift correction, (c) elliptical calibration using an aluminum standard, (d) ascertainment of lattice vectors and angles by random transform of the Bragg vector map, calculate the vector lengths and angles for all diffraction patterns, (e) strain maps generated from the dataset, and (f) rotational calibration for real space analysis.

Further description of the py4DSTEM workflow please reference Savitzky et al.[1]

### Particle selection and optimization of multiple instrumentation use

For microscopy evaluation, $Li_xFePO_4$ platelets were ultrasonically agitated in ethanol and then drop casted onto copper TEM grids. The particles were stored in a low vacuum desiccator to prevent air and moisture exposure.

To ensure ease of sample acquisition, mapping, and transfer between microscopy instruments, a Ted Pella formvar, carbon- type B index grid was used. Grid selection is highly important as detecting features within the grid bars ease identification of particles when the sample is imaged at a multitude of magnifications and resolutions using the STEM, SEM, an optical microscope, and the Advanced Light Source. Ensuring the grid is inserted right-side up with the notch at the twelve o'clock position also facilitates ease when imaging between multiple locations and set-ups (Figure 3S).

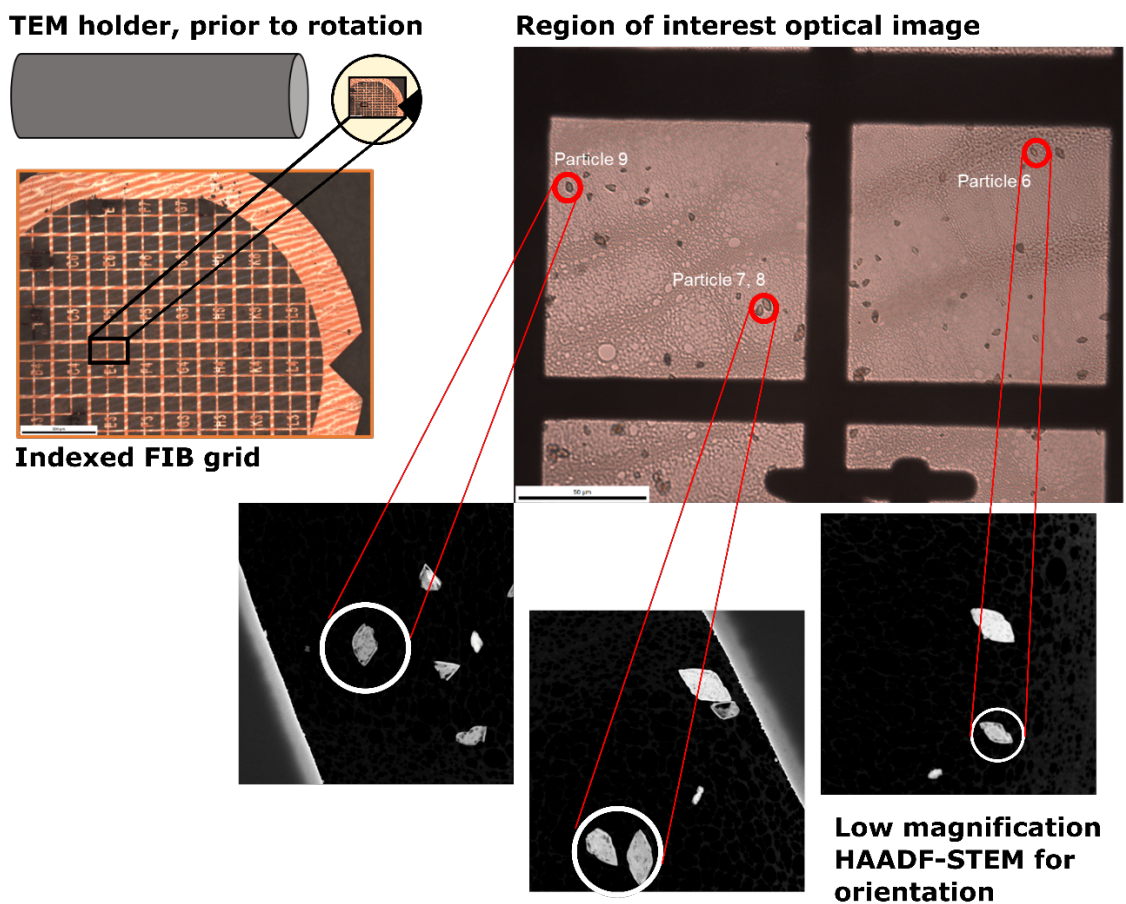

**Figure 3S.** Schematic of TEM holder prior to insertion, the formvar index TEM grid is placed right-side up with the notch at the 12 o'clock position. Low magnification imaging via the optical microscope and HAADF-STEM allows for mapping of particle placement for multi-instrument use.

While the sample stage for the electron microscope and the stage for the Advanced Light Source accommodates 3 mm typical copper TEM grids, the ALS has greater restrictions in stage movement. This stage restriction necessitates checking, with the optical microscope, prior to 4D-STEM acquisition that a moderate density of particles is well dispersed within a four to six grid region.

For best results, select particles away from grid bars to avoid shadowing effects which reduce the signal to noise ratio during the 4D-STEM acquisition. 4D-STEM should be done prior to any X-ray techniques as this electron microscopy technique is more sensitive to particle thickness, density, and isolation from other particle fragments than the X-ray techniques.

During the 4D-STEM data and high magnification HAADF-STEM image acquisition of the particles, low magnification images are also collected to develop a montage map of the grid region of interest. Using this STEM low magnification map, optical images are then acquired of this same region at 10x, 20x, and 50x. The grid is imaged facing up and facing down to ensure the next sample iteration or user is able to orient themselves regardless of sample placement or stage orientation to the collector (Figure 3S). SEM images are also acquired for grid orientation as well as to ascertain how the surface morphology compares to the electron microscopy data, i.e., the HAADF-STEM images (Figure 4S). With detailed orientation mapping, the same particles can be imaged within a variety of instrumentation.

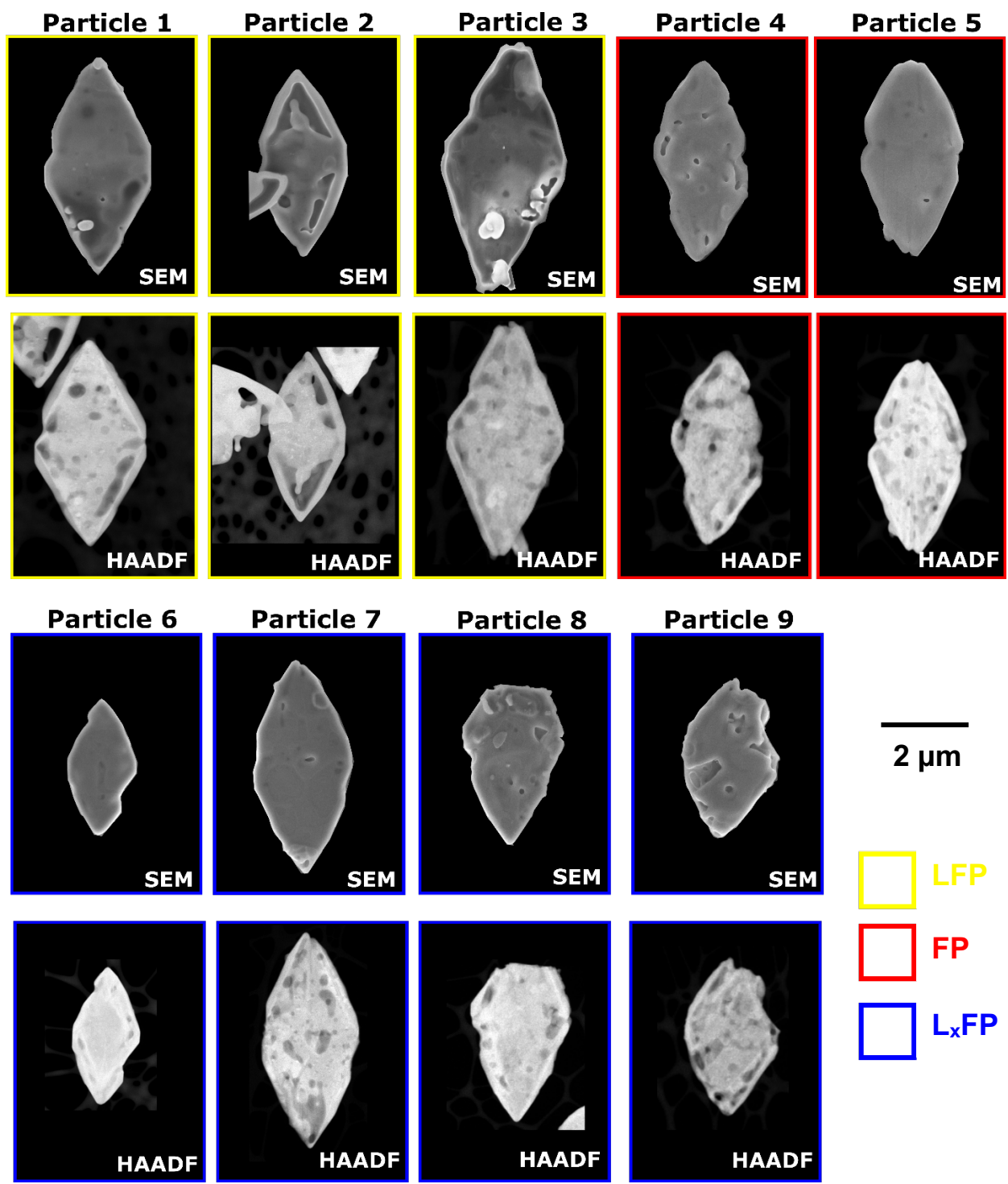

**Figure 4S.** SEM and HAADF-STEM images acquired of Particles 1-3 (LFP), 4-5 (FP), and 6-9 ($L_xFP$). While HAADF-STEM imaging shows void proliferation throughout the platelets, SEM indicates the majority of these voids are internal.

### X-ray microscopy characterization

STXM measurements were performed at beamlines 7.0.1.2 (COSMIC) at ALS using a zone plate with outer zone width of 45 nm. Fe $L_3$ edge was used to determine the local state of charge. LFP and FP were chosen as reference samples, with a full range energy stack taken at a step size of 0.2eV from 696 eV to 740 eV. Local composition was thus determined based on the absorption at 5 selected energies using a non-negative least square optimization algorithm. The STXM generated a 30 nm resolution.

Ptychography measurements were performed at beamline 7.0.1.2 (COSMIC) at ALS and were taken in double exposure mode. The same particle was imaged in STXM before and after x-ray ptychography measurement and no major changes in local composition were observed, suggesting low x-ray damage in terms of composition. Ptychography measurements yield a resolution of 10 nm, calculated based on the Fourier ring correlation method with half bit threshold.

### Pixel-by-pixel alignment of X-ray and 4D-STEM datasets

The 4D-STEM experiments and x-ray ptychography datasets cover roughly the same field of view for each particle, but without pixel-by-pixel correspondence. Additionally, because both methods use a scanning probe where the pixels are recorded serially, there are slight image distortions in all scans due to sample drift (including possibly between different X-ray energy channel measurements). Therefore, we used custom Matlab code to align all images from the different measurement channels.

We assumed the distortions / misalignments were purely affine, meaning alignment of a measured image to another reference image depends only on translation shifts in x and y, and the 2x2 linear coordinate transformation matrix that maps one coordinate system into another. For each pair of images, one was defined as the reference image, and the other was distorted to maximize a scoring function defined as the correlation of all pixels inside a mask 4 pixels inset from the particle boundaries. At each iteration, the transformation matrix was updated using gradient descent, and the translation was determined from cross correlation. In all cases the alignments converged in ~100 iterations at the most as long as reasonable initial guess was provided for the alignments.

The x-ray channels were initially aligned to each other, and then composition was estimated using the method described elsewhere in Methods. Next, the x-ray channels and STEM-HAADF images were aligned using the 4D-STEM measurement as a reference. For the x-ray measurements, the correlation between composition and the local c/a was maximized, as these were the two measurement channels we determined had the highest initial correlation. The HAADF-STEM images were aligned to a virtual dark field image created from the 4D-STEM dataset for the same reason.

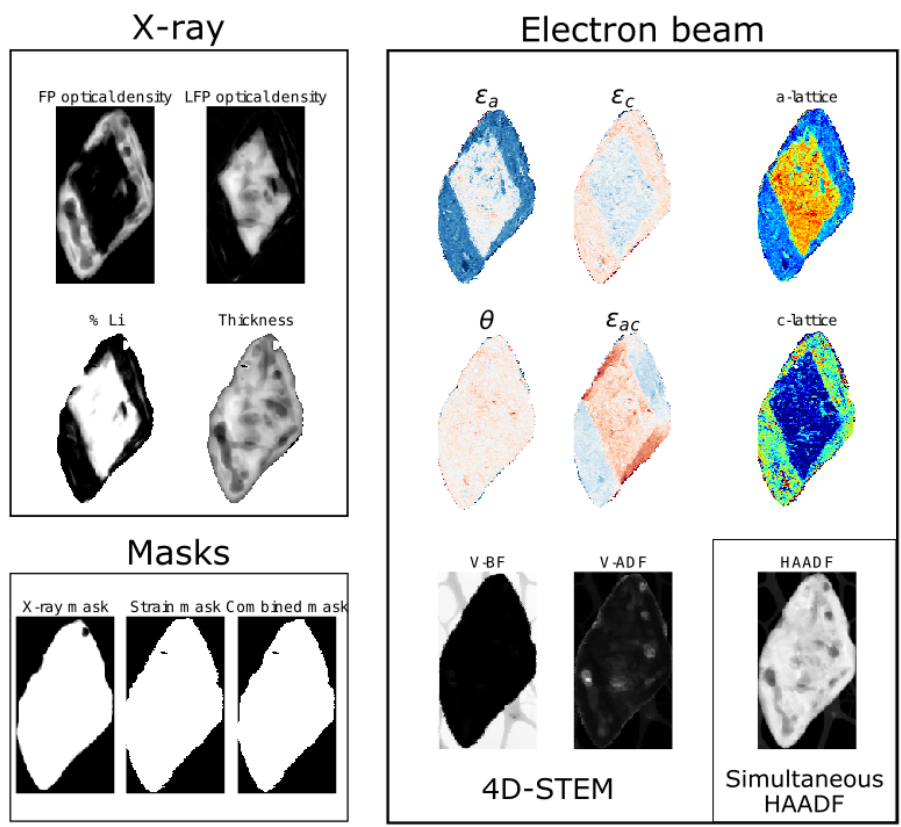

**Figure 5S.** All information channels, after pixel-by-pixel alignment. The FP/LFP optical densities were computed from Fe $L_3$ edge in the x-ray data, and the percent lithiation were in turn derived from these. The virtual bright-field (V-BF) and virtual annular dark-field (V-ADF) were computed by integrating over a selected region of diffraction space in the 4D-STEM data. The infinitesimal strain matrices and a- and c- lattice parameters were computed by detecting Bragg scattering then fitting the detected peaks to a reciprocal lattice. The strain mask represents pixels where sufficiently many Bragg disks were detected to fit the lattice vectors, the x-ray mask represents pixels where there existed sufficient signal to align with the 4D-STEM.

### Custom colormaps for *a*, *c*, and %Li

The colormaps used to show the lattice parameters and percent lithiation in Figures 1 and 3 use a custom colormap designed to make immediately visually apparent the relationship between the value of a given parameter at some position on the image and the expected value of this parameter associated with each lithiation phase.

Figures 6S and 7S shows how this was done for the a-lattice parameter. Firstly, the distribution of the parameter over the entire population of particles was calculated. Secondly, gaussians were fit to the two lobes of the distribution; we then define 5 control points. The first three control points are the means of the two gaussians and the midpoint of the two means. The final two control points, which will be the minimum and maximum values of the color-scale, are equally spaced below and above the two gaussian means. To ensure fully lithiated, fully delithiated, and intermediate states are readily visible, we set the mean associated with FP to a bright red, the mean associated with LFP to a bright yellow, and the midpoint to blue. This last choice was found to be helpful in identifying boundary regions and behavior. The minimum and maximum values are set to taper off in saturation from the closest FP / LFP control point, while leaving the hue unaltered, to indicate a region that is well within the FP or LFP domain, but farther into the distribution tails.

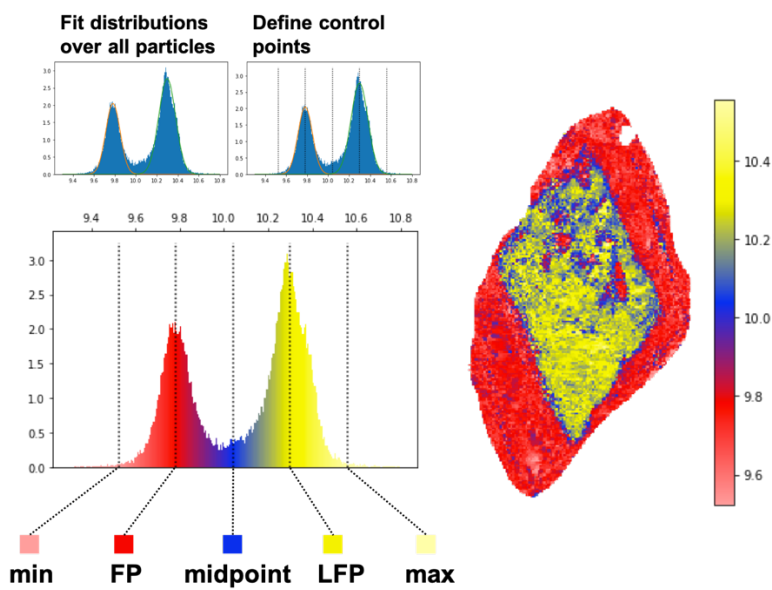

**Figure 6S.** Definition of a custom colormap for the *a*-lattice parameter, fixing the values associated with 'pure' FP/LFP and intermediate states to be immediate visually apparent.

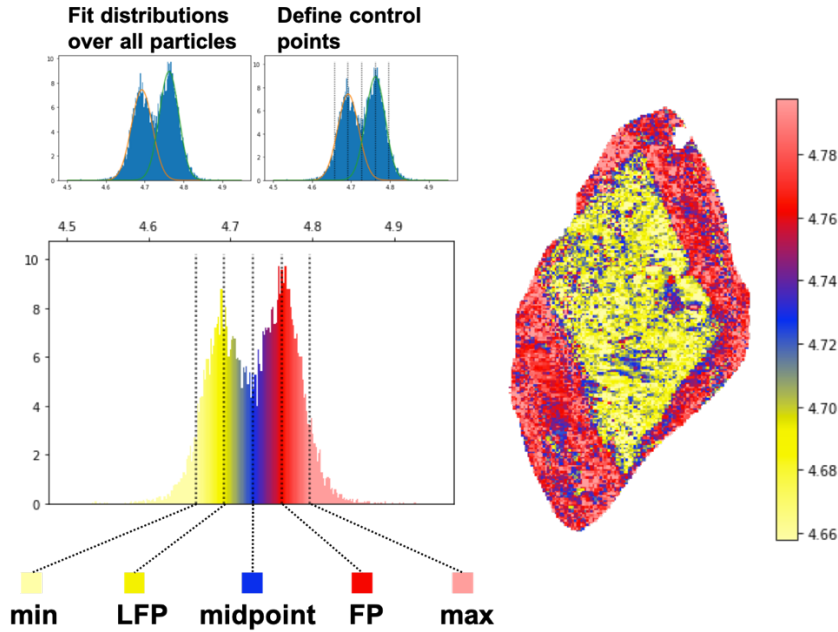

**Figure 7S.** Definition of a custom colormap for the $c$-lattice parameter. Compared to the $a$-lattice parameter, the two gaussians corresponding to distributions of 'pure' FP / LFP are overlapping, leading to the much noisier appearance of the $c$-lattice parameter map.

An important advantage of this approach is that it allows us to inspect maps of the various parameters and easily compare them to one another, and to their expected and observed relationships to the lithiation phase. The lithiation maps used a simpler implementation of the same concept, wherein three control points were used, 0, 50, and 100 percent lithiation, and identified with red, blue, and yellow.

### $L_xFP$ statistics

Means, medians, and standard deviations were calculated directly for the FP and LFP particles. For the $L_xFP$ particles, these quantities were also determined separately for each of the FP and LFP dominant regions by creating masks for each region type using the x-ray optical density channel, then computing the distributions for the $a$- and $c$-lattice parameters from all five particles, as well as computing their means, medians and standard deviations. Note, the means and medians differ meaningfully, especially in the $a$-lattice parameter, due to the skewness of the distributions. See Figure 8S and 9S.

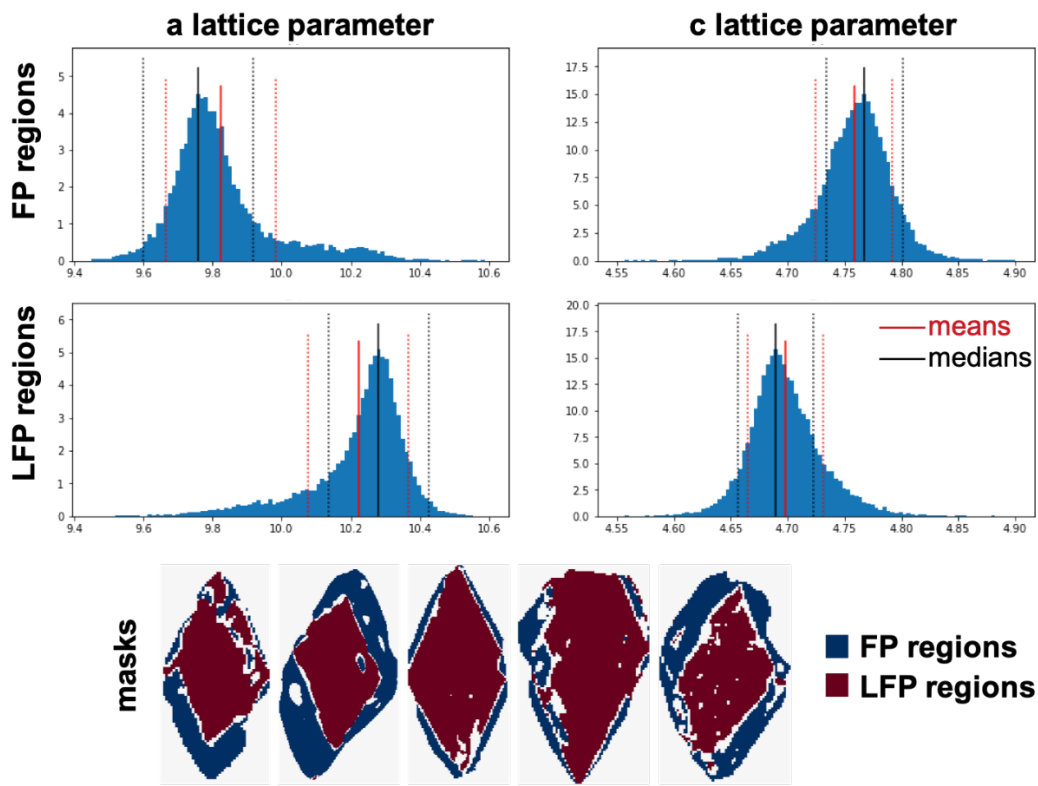

**Figure 8S.** Distributions of the *a*- and *c*-lattice parameters for the FP and LFP regions of the L$_x$FP particles. Segmentation masks are shown below.

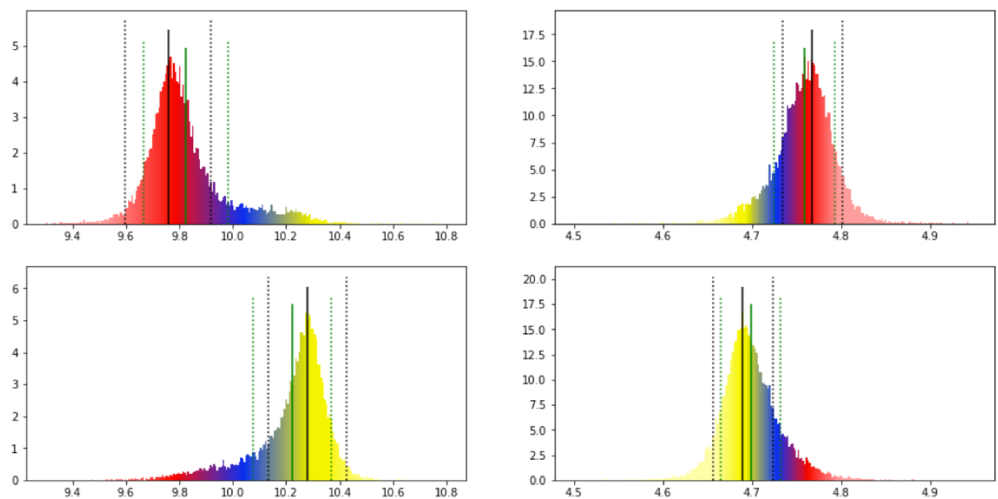

**Figure 9S.** Distributions of the *a*- and *c*-lattice parameters for the FP and LFP regions of the L$_x$FP particles shown with the custom colormaps described above.

**Chemical expansion coefficients**

We computed the means and 5/25/75/95-percentiles for the a- and c- lattice parameters versus % Li for each particle measured. Figure S10 shows the results. Figure 2 from the main paper shows similar results for the a-lattice and for each stage of lithiation, i.e. combining all LFP particles, all $L_xFP$ particles, and all FP particles.

Here we additionally compute linear fits to the lattice parameter vs. lithiation data for each particle. The slope of each fit is the best linear approximation to the chemical expansion coefficient, representing the change in structure (lattice parameter) per unit change in chemistry (lithiation). The coefficients derived for each particle were used to tabulate the values in Table 1 of the main text.

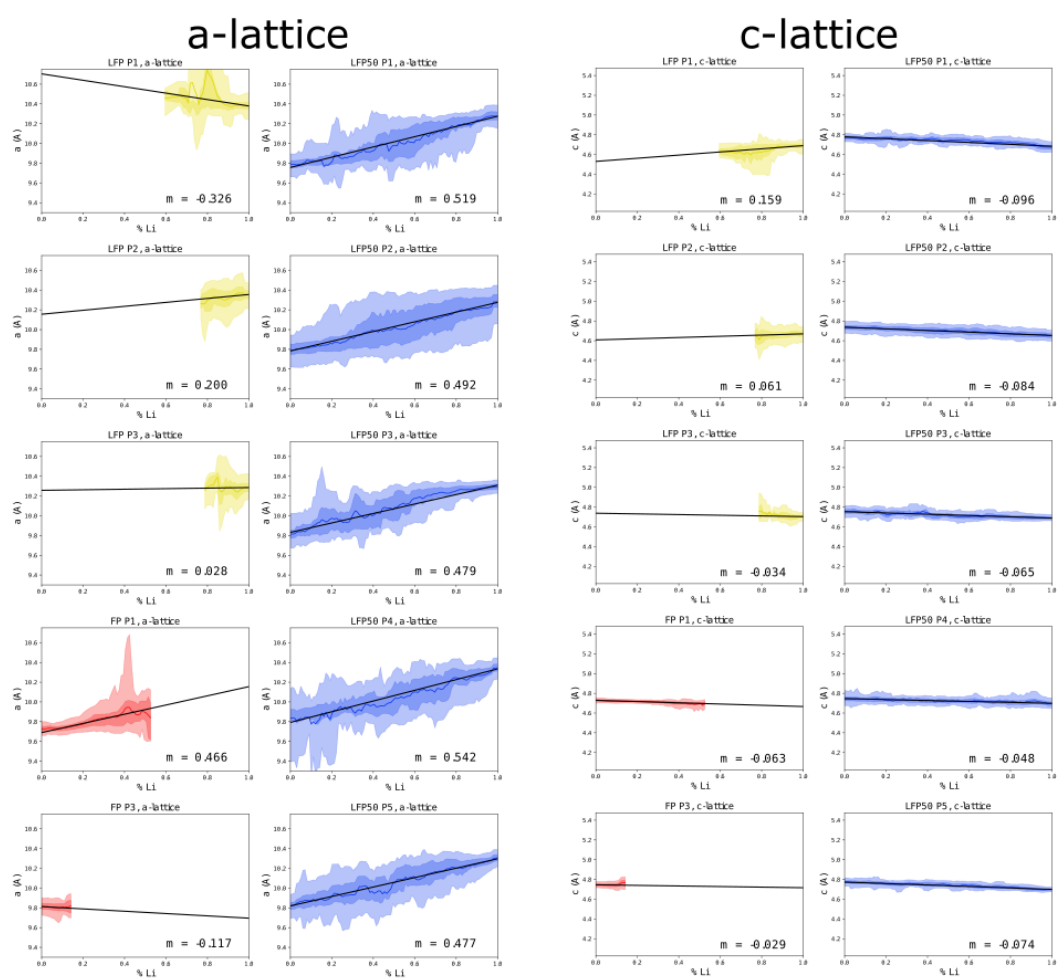

**Figure 10S.** Means, percentiles, and linear fits to the a- and c- lattice parameters vs. lithiation for all 10 particles.

**Identification of anomalous region with non-linear a-lattice behavior**

The basic premise of this approach is to calculate the deviation from Vegard's law at each point on the sample. This analysis was performed on the high pixel density scan of particle 10.

First, we compute the mean values of $a$ wherever the percent lithiation is 0% and 100%. Call these $a_{FP}$ and $a_{LFP}$, respectively. Vegard's law predicts that for some other lithiation value $L$, the local value of $a$ will obey.

$a = a_{FP} + L(a_{LFP} - a_{FP})$

We therefore compute the deviation from Vegard's law $\delta$, given by:

$\delta = a - a_{FP} - L(a_{LFP} - a_{FP})$

When $\delta > 0$, then $a$ is greater than expected for this $L$, and vice versa.

Figure 11S shows the resulting plot. We find an area with anomalously low values of $a$, shown in darker red in the figure. We segmented these regions by first thresholding, then performing a morphological closing to remove single pixels. To be conservative in what we deemed should robustly be included as 'anomalous', we then performed a 1-pixel binary erosion, then excluded 4-connected regions of fewer than 10 pixels.

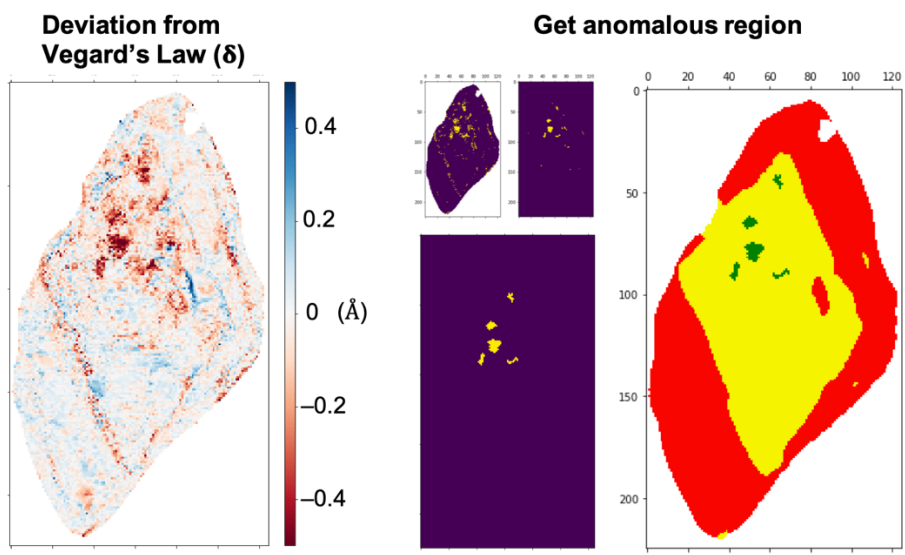

**Figure 11S.** Computation of the anomalous regions in $L_xFP$ particle 2, using a metric of deviation from Vegard's law.

The resulting masks were then used as discussed in Figure 4 and the body of the main text. We note the deviation from Vegard's law captured and segmented here can also be observed by direct comparison of the lithiation and $a$-lattice parameter plots of this particle, shown below in Figure 12S.

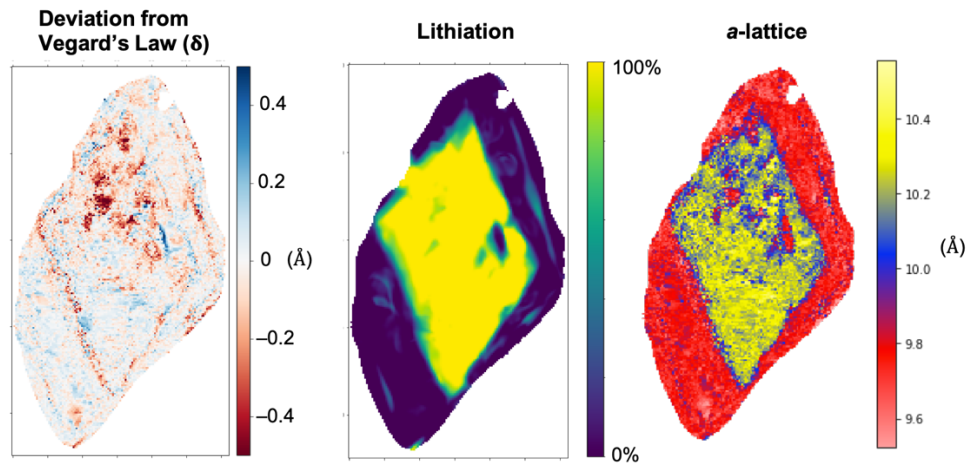

**Figure 12S.** Deviation from Vegard's law in the anomalous regions identified above are also visible by direct comparison of the %Li plot and the a-lattice plot. In the lithiation plot, the upper middle-left area of the lithiated core is nearly constant, while in the *a* plot there are many small islands of red in this area, indicated a lattice parameter consistent with FP, rather than LFP.

**Additional mapping possible with 4D-STEM analysis**

Pixel-by-pixel correlation was performed on all imaging techniques available via 4D-STEM and X-ray microscopy. This correlation allows for direct comparison of HAADF-STEM, V-ADF, V-BF, optical density of FP and LFP, as well as thickness images (Figure 13S) for all ten particles.

Additional to lattice parameter mapping, 4D-STEM datasets allow for the calculation of both shear and rotation angle changes within LFP platelets. Figure 14S shows an example of shear and rotation for LFP (Particle 1), FP (Particle 5), and $L_xFP$ (Particle 6). For all particles, there is a minimal change in shear as the particles are delithiated. For the rotational angle, with the fully delithiated and lithiated particles, the rotation is approximately zero. For the 50% percent delithiated LFP particles there is a rotational shift within the delithiated regions from -2 to 2 degrees. This shift is related to the orientation of the lithiated and delithiated regions with respect to one another.

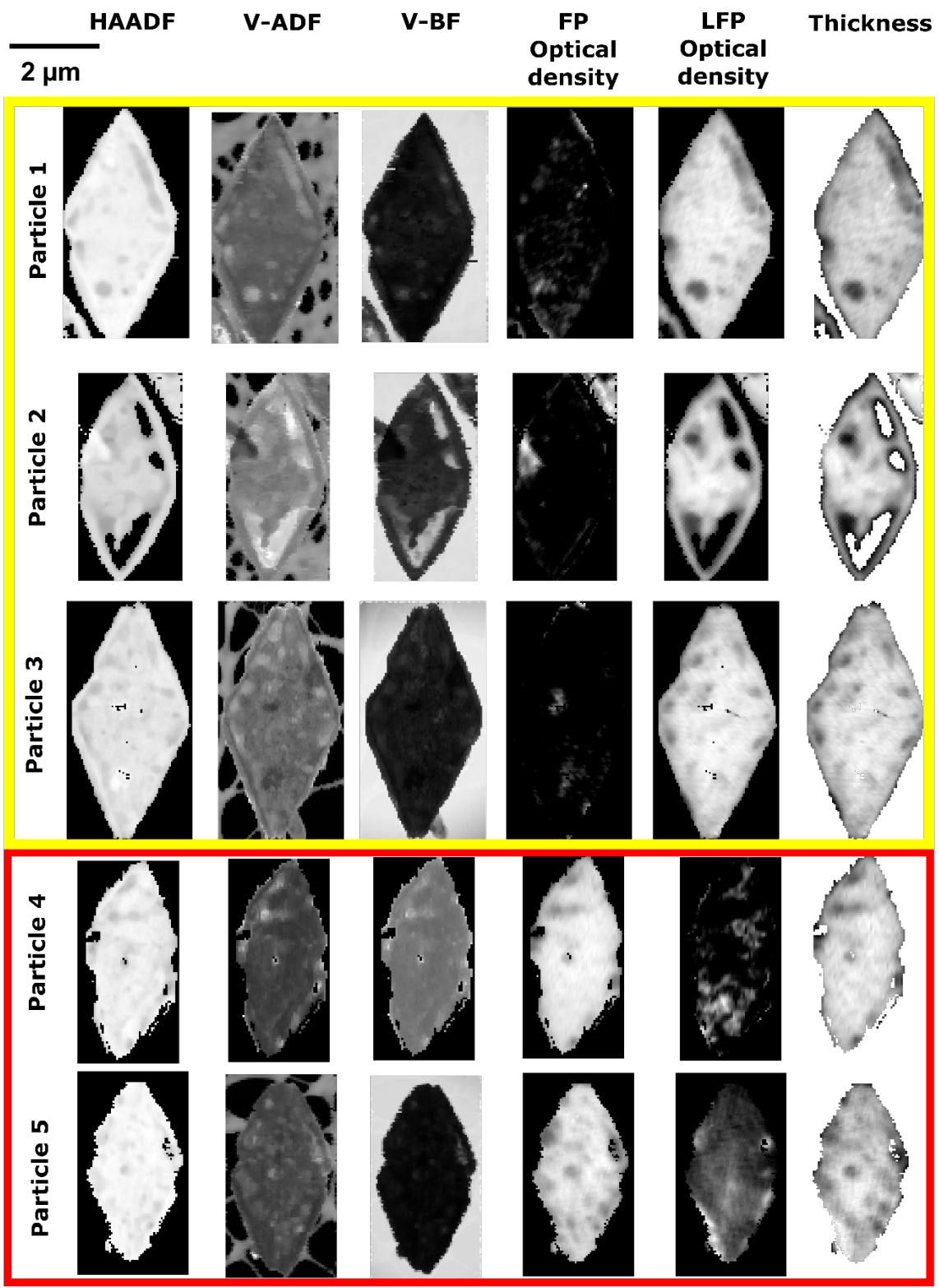

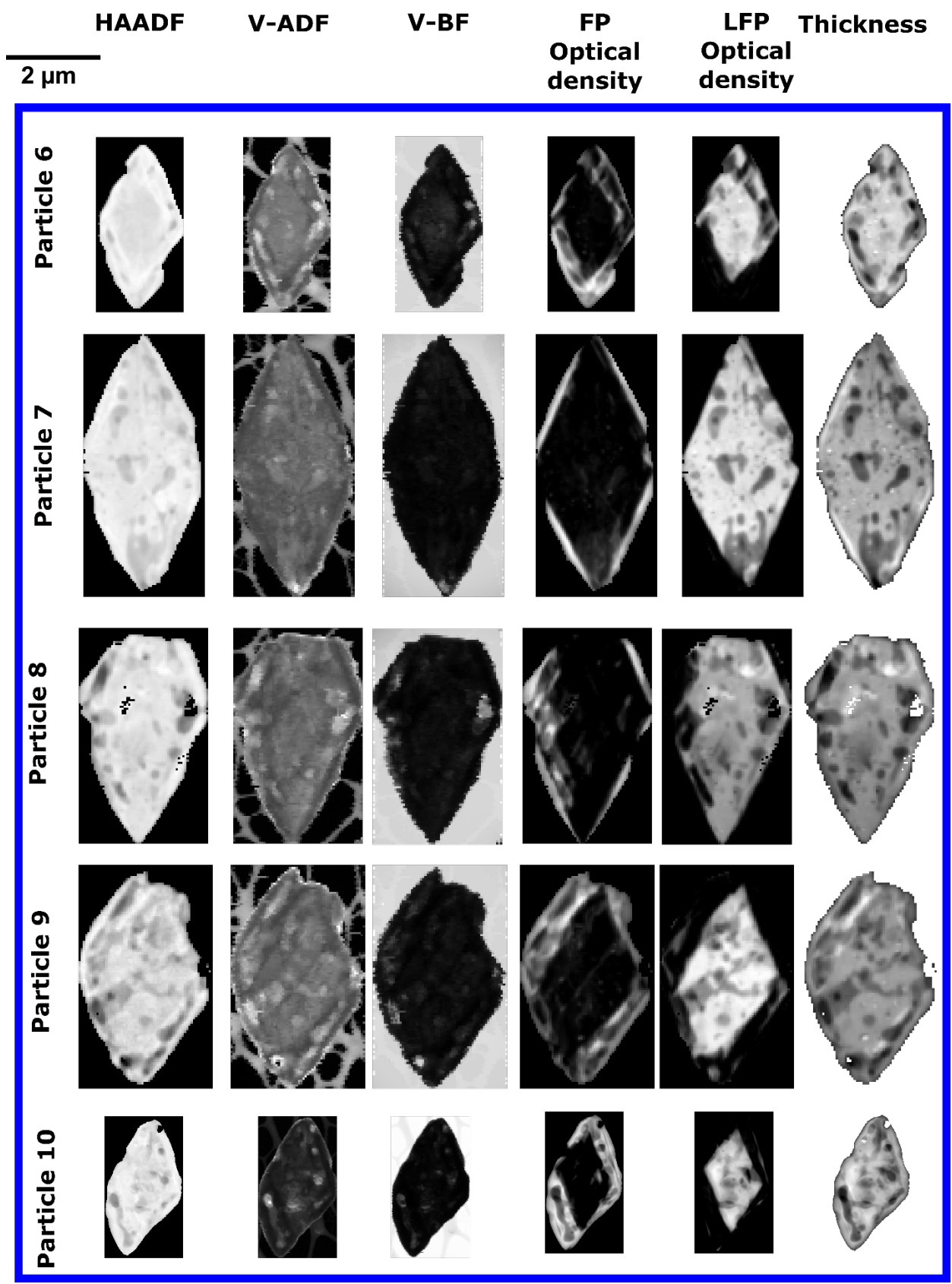

**Figure 13S.** Pixel-by-pixel correlation of HAADF-STEM, V-ADF, V-BF, FP and LFP optical density as well as thickness maps for all particles.

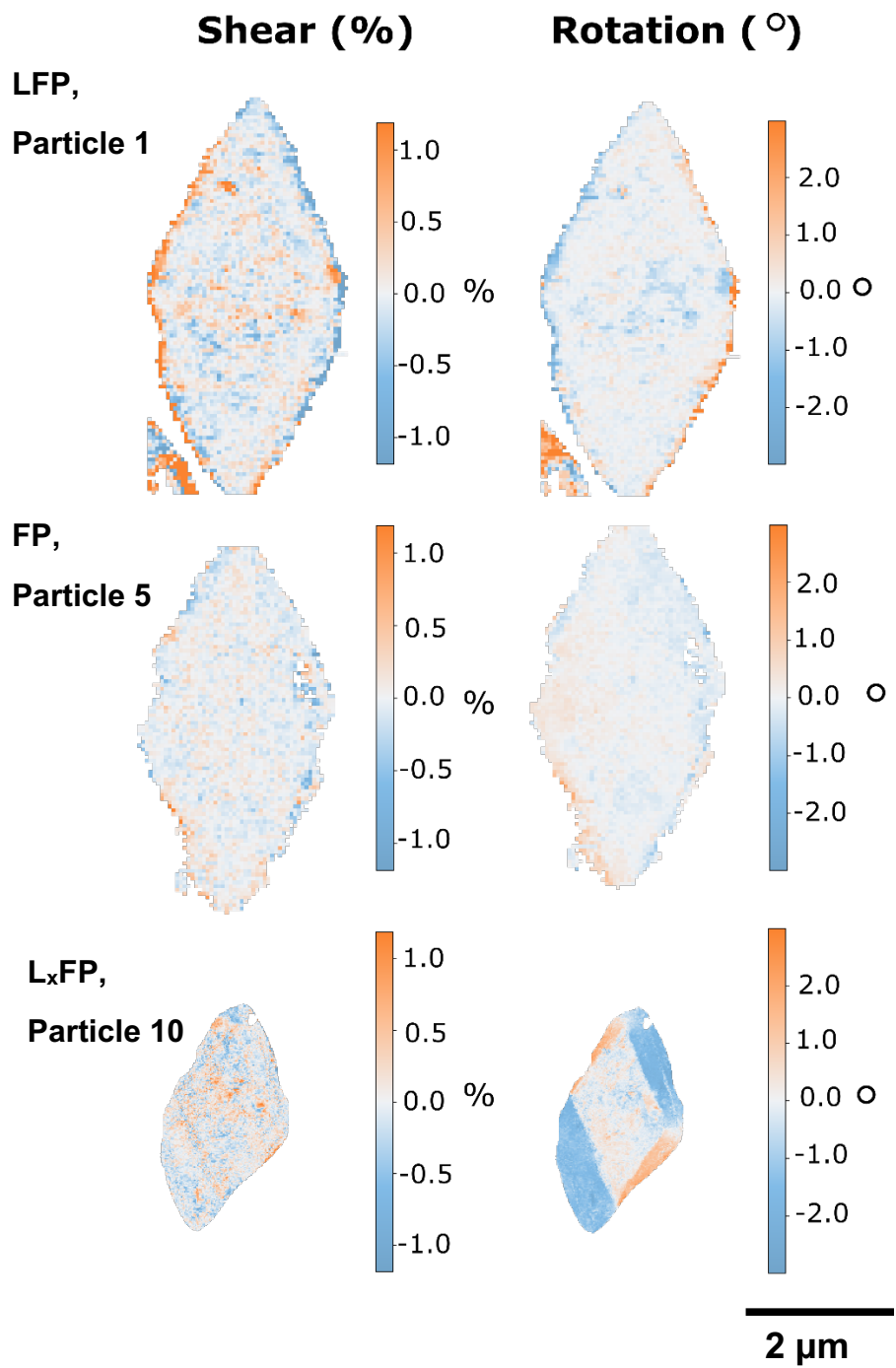

**Figure 14S.** Shear (%) and rotation angle (°) maps of LFP (Particle 1), FP (Particle 5), and L$_x$FP (Particle 6).

## Cross validation RMSE calculation

Cross validation was used to compute the root mean square error of the lattice vector fits, which were used to compute the strain. At each diffraction pattern, the detected Bragg peaks were divided into two groups of equal size – a training group, and a test group. The lattice vectors were fit using the training group, then, the positions of the test group peaks were predicted using these fits. We then computed the root mean square of the difference between these predictions and the measured test group positions. This procedure was repeated 5 times for each scan position, dividing the peaks randomly into different training and test groups each time. The results are shown in Fig. 4 in the main text, and in Fig. 15S below.

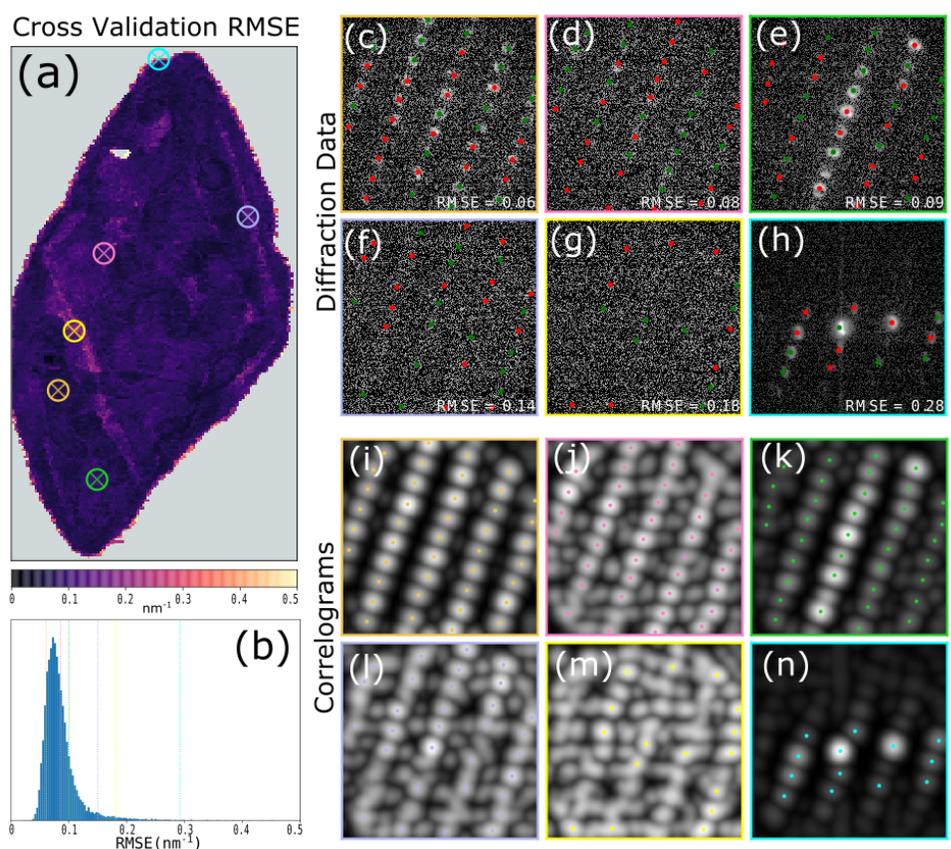

**Figure 15S.** Cross validation computation of the RMSE of the lattice vector fits. (a) The RMSE at each scan position, and (b) the histogram of these values. (c-h) The detected peaks, divided into training (green) and test (red) data, in four sample diffraction patterns. The diffraction patterns in (c-e) are representative of much of the data and have comparatively low RMSE values. The next three patterns represent regions of higher error. In (f) and (g), the high error appears to follow from a low signal-to-noise ratio, which may result from increased thickness of the particle, or tilt of the crystal lattice. In (h) the higher RMSE results from significant redistribution of intensity within the central disk, likely due to the rapid change in potential at the edge of the particle. High CV RMSE also results in cases in which nearly all of the detected Bragg peaks are co-linear, such that the non-colinear test peaks cannot be predicted from the training peaks.

The root-mean-square error associated with the reciprocal lattice vectors identified at each scan position was assessed in two ways: with cross validation, and from the error associated with fitting the vectors to a lattice formed by the detected Bragg peak positions. These approaches are shown in Fig. 16S.

We find that the two RMSEs are in near agreement, with the CV error slightly higher, for most pixels where the error is small. This corresponds to diffraction patterns with relatively high signal-to-noise ratios, like those shown in Fig. 15S (c-e). However, for diffraction patterns with low SNR (Fig. 15S f, g), or those with any source of systematic error (Fig. 15S h), the lattice fit error is a massive underestimate of our certainty of the correctness of the measured lattice vectors at these pixels. This is the primary reason computing the error with cross validation, rather than simply using the lattice fit error, is important. In Fig. 16S d, this is represented by the long tail of counts with high CV RMSE, which are missed entirely when using only the lattice fitting error.

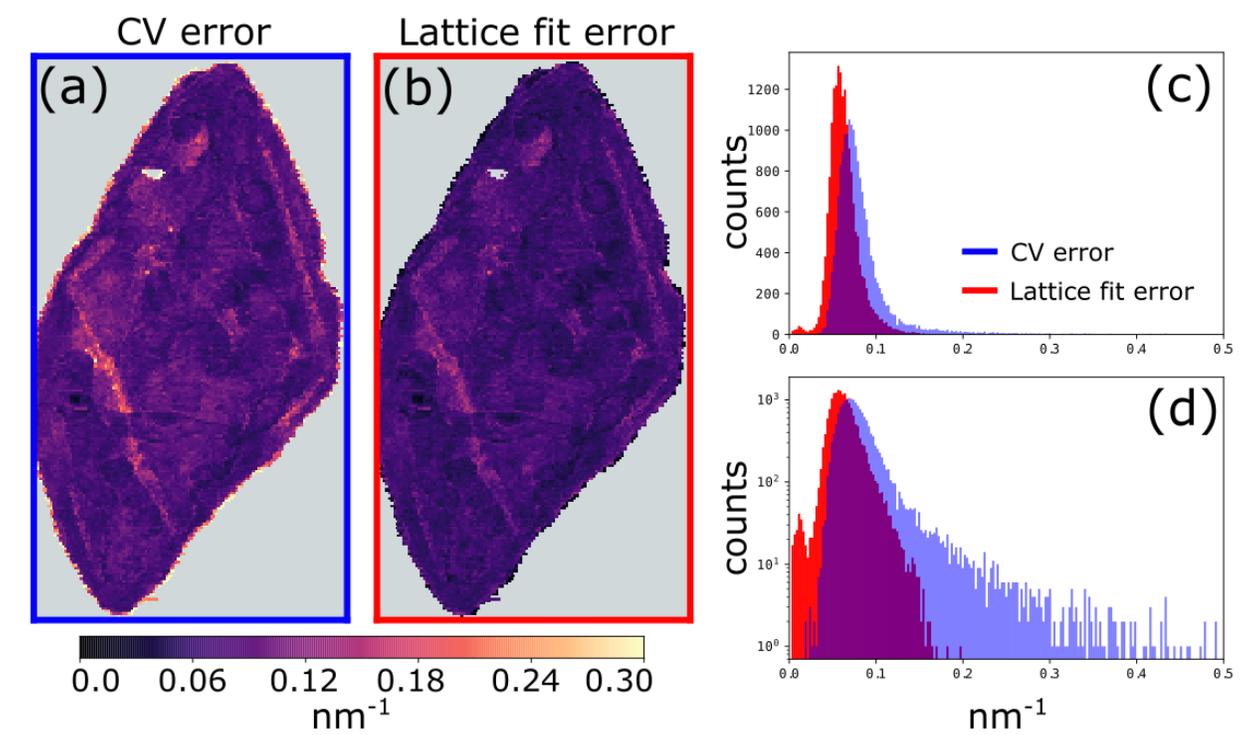

**Figure 16S.** Comparison of CV and lattice fitting RMS error in particle 10. (a, b) The RMSE associated with the reciprocal lattice vectors at each scan position, computed by cross validation (a) and from the error associated with the lattice vector fitting (b). (c, d) The histograms of these values, plotted with a linear (c) and logarithmic (d) scale.

**Systematic error**

The thickness of these samples makes systematic error due to strong scattering effects in the electron beam inevitable, leading to the possibility that slight mistilts, for instance, may affect measurements of the lattice. We assess that systematic error in Fig. 17S. Here, we computed and plotted the lattice parameter distributions for both the a- and c- lattice for all 10 particles. In each plot, the literature values for the lattice parameters are shown as dashed red(yellow) lines for FP(LFP). We fit gaussians to each peak in the distributions, and then assess the systematic error as the deviation of the means of gaussian fits from the literature values. For comparison, we also compute the random error as the standard deviation of the gaussians. Over all particles, we find a systematic error associated with the a-lattice of 3.6 ± 1.5 pm, and a systematic error associated with the c-lattice of 2.2 ± 1.8 pm.

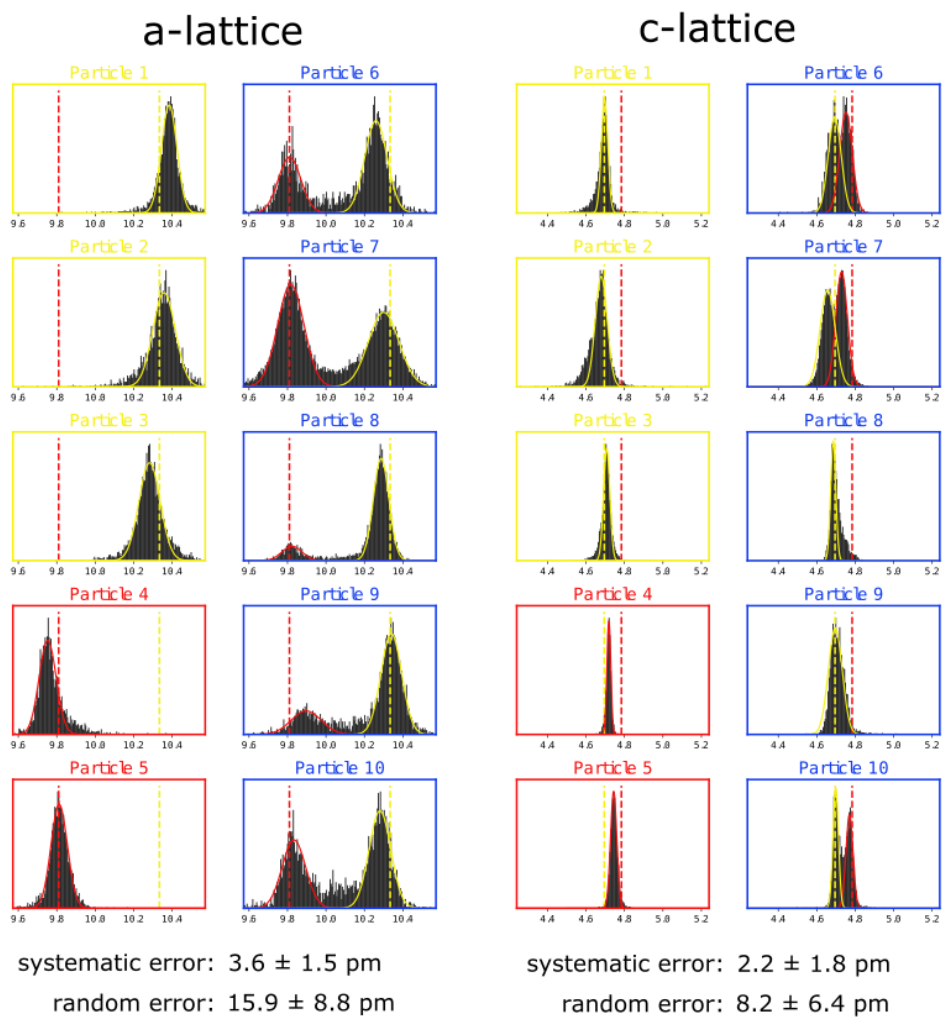

**Figure 17S.** Assessment of the systematic error in lattice parameter measurements.